\documentclass[12pt]{iopart}

\usepackage{hyperref} 
\usepackage{graphicx}
\usepackage{color}
\usepackage{dcolumn}
\usepackage{bm}
\usepackage{verbatim} 
\usepackage{cite} 
\usepackage[normalem]{ulem}

\begin{document}


    \long\def \cblu#1{\color{blue}#1}
    \long\def \cred#1{\color{red}#1}
    \long\def \cgre#1{\color{green}#1}
    \long\def \cpur#1{\color{purple}#1}

\newcommand{\eric}[1]{{\color{blue}#1}}
\newcommand{\guido}[1]{{\color{violet}#1}}
\newcommand{\matthias}[1]{{\color{blue}#1}}
\newcommand{\fabian}[1]{{\color{blue}#1}}
\newcommand{\di}[1]{{\color{blue}#1}}
\newcommand{\ericC}[1]{{\color{red}\textit{\textbf{Eric:} #1}}}
\newcommand{\guidoC}[1]{{\color{red}\textit{\textbf{Guido:} #1}}}
\newcommand{\matthiasC}[1]{{\color{red}\textit{\textbf{Matthias:} #1}}}
\newcommand{\fabianC}[1]{{\color{red}\textit{\textbf{Fabian:} #1}}}
\newcommand{\diC}[1]{{\color{red}\textit{\textbf{Di:} #1}}}

\def\FileRef{
\def\FName{/u/dihu/Draft/RunawayDrift/POP\_pub/referee/List.d}

{
\newcount\hours
\newcount\minutes
\newcount\min
\hours=\time
\divide\hours by 60
\min=\hours
\multiply\min by 60
\minutes=\time
\
\advance\minutes by -\min
{\small\rm\em\the\month/\the\day/\the\year\ \the\hours:\the\minutes
\hskip0.125in{\tt\FName}
}
}}

\mathchardef\muchg="321D
\let\na=\nabla
\let\pa=\partial

\let\muchg=\gg

\let\t=\tilde
\let\ga=\alpha
\let\gb=\beta
\let\gc=\chi
\let\gd=\delta
\let\gD=\Delta
\let\ge=\epsilon
\let\gf=\varphi
\let\gg=\gamma
\let\gh=\eta
\let\gj=\phi
\let\gF=\Phi
\let\gk=\kappa
\let\gl=\lambda
\let\gL=\Lambda
\let\gm=\mu
\let\gn=\nu
\let\gp=\pi
\let\gq=\theta
\let\gr=\rho
\let\gs=\sigma
\let\gt=\tau
\let\gw=\omega
\let\gW=\Omega
\let\gx=\xi
\let\gy=\psi
\let\gY=\Psi
\let\gz=\zeta

\let\lbq=\label
\let\rfq=\ref
\let\na=\nabla
\def\daI{{\dot{I}}}
\def\dsq{{\dot{q}}}
\def\dgj{{\dot{\phi}}}

\def\bgs{\bar{\sigma}}
\def\bgh{\bar{\eta}}
\def\bgg{\bar{\gamma}}
\def\bgy{\bar{\psi}}
\def\bgF{\bar{\Phi}}
\def\bgY{\bar{\Psi}}

\def\baF{\bar{F}}
\def\bsj{\bar{j}}
\def\baJ{\bar{J}}
\def\bsp{\bar{p}}
\def\baP{\bar{P}}
\def\bsx{\bar{x}}

\def\hgj{\hat{\phi}}
\def\hgq{\hat{\theta}}

\def\HaT{\hat{T}}
\def\HaR{\hat{R}}
\def\Hsb{\hat{b}}
\def\Hsh{\hat{h}}
\def\Hsz{\hat{z}}

\let\gG=\Gamma
\def\taA{{\tilde{A}}}
\def\taB{{\tilde{B}}}
\def\taG{{\tilde{G}}}
\def\tsp{{\tilde{p}}}
\def\tsv{{\tilde{v}}}
\def\tgF{{\tilde{\Phi}}}

\def\wgx{{\bm{\xi}}}
\def\wgz{{\bm{\zeta}}}

\def\wse{{\bf e}}
\def\wsk{{\bf k}}
\def\wsi{{\bf i}}
\def\wsj{{\bf j}}
\def\wsl{{\bf l}}
\def\wsn{{\bf n}}
\def\wsp{{\bf p}}
\def\wsr{{\bf r}}
\def\wsu{{\bf u}}
\def\wsv{{\bf v}}
\def\wsx{{\bf x}}

\def\vaB{\vec{B}}
\def\vse{\vec{e}}
\def\vsh{\vec{h}}
\def\vsl{\vec{l}}
\def\vsv{\vec{v}}
\def\vgn{\vec{\nu}}
\def\vgk{\vec{\kappa}}
\def\vgt{\vec{\gt}}
\def\vgx{\vec{\xi}}
\def\vgz{\vec{\zeta}}

\def\waA{{\bf A}}
\def\waB{{\bf B}}
\def\waD{{\bf D}}
\def\waE{{\bf E}}
\def\waF{{\bf F}}
\def\waJ{{\bf J}}
\def\waV{{\bf V}}
\def\waX{{\bf X}}

\def\R#1#2{\frac{#1}{#2}}
\def\btbl{\begin{tabular}}
\def\etbl{\end{tabular}}
\def\bqbl{\begin{eqnarray}}
\def\eqbl{\end{eqnarray}}
\def\ebox#1{
  \begin{eqnarray}
    #1
\end{eqnarray}}


\def \cred#1{{\color{red}\sout{(#1)}}}
\def \cblu#1{{\color{blue}#1}}
\title[Heat flux \& impact during SPI]{Plasmoid drift and first wall heat deposition during ITER H-mode dual-SPIs in JOREK simulations}
\author{D. Hu$^{1,*}$, F. J. Artola$^2$, E. Nardon$^{3}$, M. Lehnen$^2$, M. Kong$^4$, D. Bonfiglio$^5$, M. Hoelzl$^6$, G.T.A. Huijsmans$^{3,7}$ \& JOREK Team\footnote{See Hoelzl et al 2021 (https://doi.org/10.1088/1741-4326/abf99f) for the JOREK Team.}}
\address{\scriptsize
$^1$Beihang University, No. 37 Xueyuan Road, Haidian District, 100191 Beijing, China.
}
\address{\scriptsize
$^2$ITER Organization, Route de Vinon sur Verdon, CS 90 046,13067 Saint Paul-lez-Durance, Cedex, France.
}
\address{\scriptsize
$^3$CEA, IRFM, F-13108 Saint-Paul-Lez-Durance, France
}
\address{\scriptsize
$^4$\'{E}cole Polytechnique F\'{e}d\'{e}rale de Lausanne (EPFL), Swiss Plasma Center (SPC), CH-1015 Lausanne, Switzerland
}
\address{\scriptsize
$^5$Consorzio RFX-CNR, ENEA, INFN, Universit\`{a} di Padova, Acciaierie Venete SpA. I-35127 Padova, Italy.
}
\address{\scriptsize
$^6$Max Planck Institute for Plasma Physics, Boltzmannstr. 2, 85748 Garching b. M., Germany
}
\address{\scriptsize
$^7$Eindhoven University of Technology, De Rondom 70 5612 AP Eindhoven, the Netherlands.
}
\ead{* hudi2@buaa.edu.cn}
\vspace{10pt}
\begin{indented}
\item[]\today
\end{indented}

\begin{abstract}
    The heat flux mitigation during the Thermal Quench (TQ) by the Shattered Pellet Injection (SPI) is one of the major elements of disruption mitigation strategy for ITER. It's efficiency greatly depends on the SPI and the target plasma parameters, and is ultimately characterised by the heat deposition on to the Plasma Facing Components (PFCs). To investigate such heat deposition, JOREK simulations of neon-mixed dual-SPIs into ITER baseline H-mode and a ``degraded H-mode'' with and without good injector synchronization are performed with focus on the first wall heat flux and its energy impact. It is found that low neon fraction SPIs into the baseline H-mode plasmas exhibit strong major radial plasmoid drift as the fragments arrive at the pedestal, accompanied by edge stochasticity. Significant density expulsion and outgoing heat flux occurs as a result, reducing the mitigation efficiency. Such drift motion could be mitigated by injecting higher neon fraction pellets', or by considering the pre-disruption confinement degradation, thus improving the radiation fraction. The radiation heat flux is found to peak in the vicinity of the fragment injection location in the early injection phase, while it relaxes later on due to parallel impurity transport. The overall radiation asymmetry could be significantly mitigated by good synchronization. Time integration of the local heat flux is carried out to provide its energy impact for wall heat damage assessment. For the baseline H-mode case with full pellet injection, melting of the stainless steel armour of the diagnostic port could occur near the injection port, which is acceptable, without any melting of the first wall tungsten tiles. For the degraded H-mode cases with quarter-pellet SPIs, which have $1/4$ total volume of a full pellet, the maximum energy impact approaches the tolerable limit of the stainless steel with un-synchronized SPIs, and stays well below such limit for the perfectly synchronized ones.
\end{abstract}

%
%
%
\maketitle
%
%

\section{Introduction}
\label{s:Intro}

The Thermal Quench (TQ) heat flux mitigation by impurity Shattered Pellet Injection (SPI) is an important part of the ITER Disruption Mitigation System (DMS) strategy, the aim of which is to evenly redistribute the thermal energy stored within the plasma onto the Plasma Facing Components (PFCs), so that localized heat deposition on the divertor or the first wall and the consequential material damage could be avoided during the TQ process \cite{Lehnen2015JNM}. The TQ mitigation efficiency thus greatly depends on both the radiated energy fraction, as well as the asymmetry of the radiation heat flux onto the first wall. 

There have been intensive studies, both experimentally \cite{Li2020NF,Sheikh2021NF,Jachmich2022NF,Jang2022FED,Stein-Lubrano2024NF} and numerically \cite{Kim2019POP,Di2021NF,Di2023NF,McClenaghan2023NF}, on both the total radiated energy and the radiation power density within the plasma volume after SPIs. It has been found in general that an apparent toroidal asymmetry in the radiation power density exists for the single SPI cases, especially in the early injection phase. The radiation power density profile tends to relax as the TQ proceeds, and the Toroidal Peaking Factor (TPF) of the radiation power density at the time of the maximum radiation power tends to be smaller compared with the maximum TPF during the whole TQ phase in both experimental observations \cite{Sheikh2021NF} and numerical simulations \cite{Di2021NF}.

Although the aforementioned studies already provide valuable insight into the asymmetric TQ mitigation process, the ultimate criterion of the disruption mitigation efficiency should be the heat flux onto the first wall and its accumulated energy impact, as the latter is directly related to the wall material temperature rise and thus the wall damage during disruption\cite{Herrmann2002PPCF}. The recent change of the ITER first wall material to tungsten offers more tolerance to such transient heat flux \cite{Barabaschi2023FEC}, however, should the accumulated energy impact exceed a certain threshold, even tungsten could experience significant melting \cite{Herrmann2002PPCF}. Meanwhile, the stainless steel cover for the port window suffers from a lower energy impact threshold which is about $1/3$ of the tungsten threshold, thus is more susceptible of melting, although such melting is acceptable as long as the melt threshold is not dramatically exceeded for the stainless steel. \cite{Klimov2013JNM,Pitts2015JNM}. Furthermore, even without exceeding the tolerable maximum energy impact, transient heat fluxes could already produce structural modification on tungsten surface which could affect the mechanical and thermal properties of the first wall \cite{Coenen2011PS,vanEden2014NF,Yuan2016NF,Wang2018NF,Yuan2019NF}. Hence it is desirable to obtain the heat flux and energy impact distribution onto the first wall after SPIs to assess the efficiency of the TQ mitigation and to validate the ITER PFC heat tolerance specifications.

To take a first look into the aforementioned heat deposition behaviour in ITER H-mode plasmas, we carry out 3D non-linear reduced MHD simulations using the JOREK code \cite{Hoelzl2021NF} and its collisional-radiative impurity module \cite{Di2021PPCF}. Two sets of equilibria are considered, a ``baseline'' H-mode and a ``degraded'' one. The latter means to take into account the natural confinement degradation in the disruption precursor phase, which significantly reduces the thermal energy before the triggering of the DMS \cite{Riccardo2005NF}. Two sets of dual-SPI configurations are also considered, one ``full pellet'' with small neon mixture ratio and one ``quarter-pellet'' with the same amount of neon atoms but one fourth the hydrogen content. For the full pellet dual-SPI into baseline H-mode case, a strong plasmoid drift along the major radial direction is found to occur due to strong local over-pressure within the newly ablated plasmoid cloud which induces polarization and subsequent $\waE\times \waB$ drift towards the Low Field Side (LFS) \cite{Rozhansky1995PPCF}. Such drift and its accompanying MHD response would undermine the assimilation when fragments are injected from the LFS mid-plane, as well as cause strong boundary heat flux as the edge becomes stochastic. These detrimental behaviours could be mitigated if the higher neon mixture ratio quarter-pellet is used so that the local over-pressure is suppressed by local radiation \cite{Matsuyama2022PRL} or by considering the degraded H-mode. Indeed, the quarter-pellet SPIs into degraded H-mode shows good impurity assimilation and radiation fraction. Moreover, the radiation heat flux onto the first wall is obtained by the Raysect/CHERAB code suite integrated within Integrated Modelling \& Analysis Suite (IMAS) \cite{CarrRaySect,CarrCHERAB} and their energy impact is calculated. It is found that the maximum energy impact for the full pellet dual-SPI into the baseline H-mode case exceeds the tolerable limit of the stainless steel near the injector, but stays well below that of the tungsten. For the quarter-pellet dual-SPI into the degraded H-mode, the maximum energy impact only exceeds the stainless steel limit slightly even when there is $1ms$ delay between the injectors, while it stays well below said limit if the synchronization is perfect.

The rest of the paper is arranged as the following: In Section \ref{s:Setup}, the two target equilibria are introduced, and our basic assumptions, simulation setup as well as the SPI configurations and parameters are presented. We also introduce our method of calculating the accumulated energy impact from a time-varying heat flux distribution onto the first wall. In Section \ref{s:PlasmoidDrift}, the plasmoid drift behaviour during H-mode SPIs and its mitigation are investigated, and the possible consequences of such drift are discussed. The radiation heat flux and its energy impact onto the first wall for both the baseline and degraded H-mode cases are shown in Section \ref{s:HeatFlux}, and their implication to the PFC damage is discussed. Finally, conclusion and further discussions on the TQ mitigation efficiency with neon SPIs are presented in Section \ref{s:Conclusion}.

\section{Simulation setup and calculation of the energy impact}
\label{s:Setup}

\subsection{Simulation setup and basic assumptions}
\label{ss:System}

\begin{figure*}
\centering
\noindent
\btbl{cc}
\parbox{2.45in}{
	\includegraphics[scale=0.27]{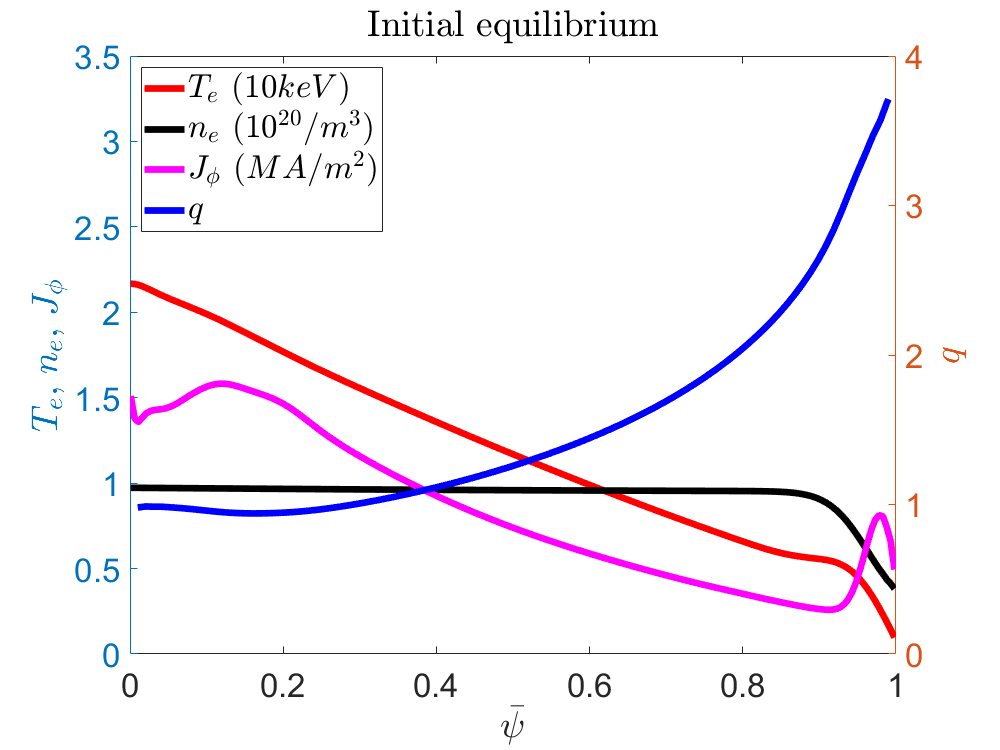}
}
&
\parbox{2.45in}{
	\includegraphics[scale=0.27]{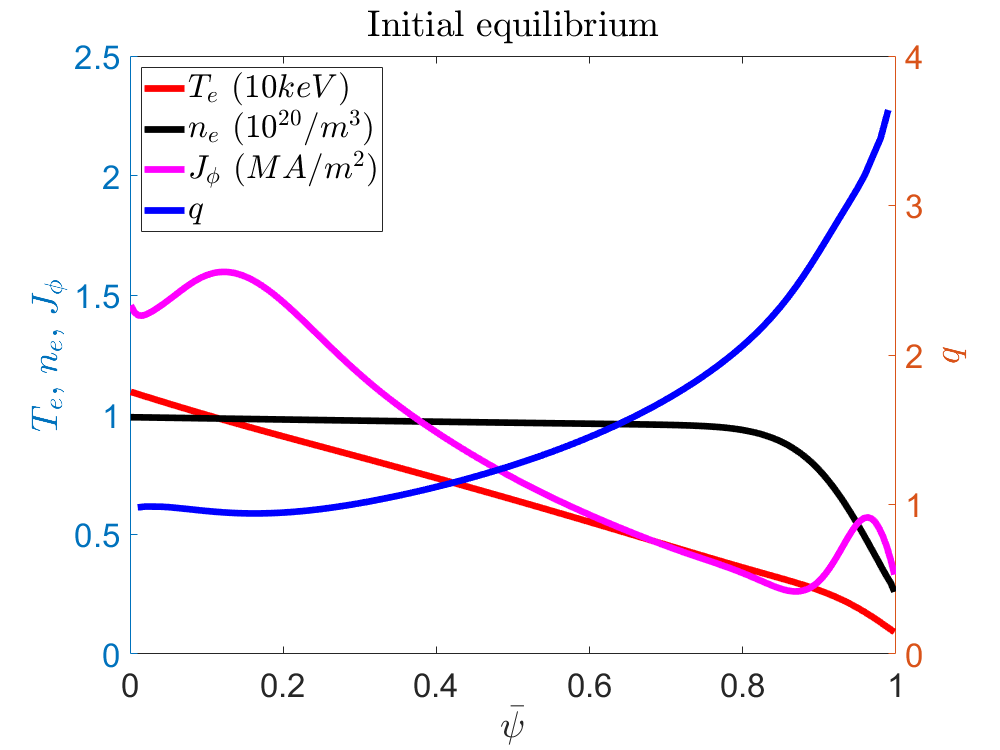}
}
\\
(a)&(b)
\etbl
\caption{The initial equilibria of (a) the baseline H-mode and (b) the degraded H-mode, $\bgy$ is the normalized poloidal flux.}
\label{fig:InitProfiles}
\end{figure*}

The 3D non-linear reduced MHD version of JOREK with collisional-radiative impurity model \cite{Hoelzl2021NF,Di2021PPCF} is used for this study, and our governing equations are the same as that described in Ref.\,\cite{Di2023NF} without modifications, so we shall not repeat them here. The target equilibria are one baseline H-mode chosen from the ITER reference scenarios \cite{Kim2018NF}, and one degraded H-mode constructed from the baseline H-mode equilibrium by increasing the perpendicular heat conduction and diffusion artificially thus producing a H-L back transition, mimicking the naturally occurring confinement degradation before the onset of the TQ \cite{Riccardo2005NF}. Both cases have the same toroidal magnetic field at $B_T=5.3T$ and the total plasma current at $I_p=15MA$. The initial equilibrium profiles, including the electron temperature, density, current density and the safety factor $q$ profiles for both target plasmas are shown in Fig.\,\ref{fig:InitProfiles}. The initial thermal energy is $190MJ$ for the degraded H-mode case and $370MJ$ for the baseline H-mode case.

\begin{figure*}
\centering
\noindent
\btbl{c}
\parbox{2.5in}{
    \includegraphics[scale=0.35]{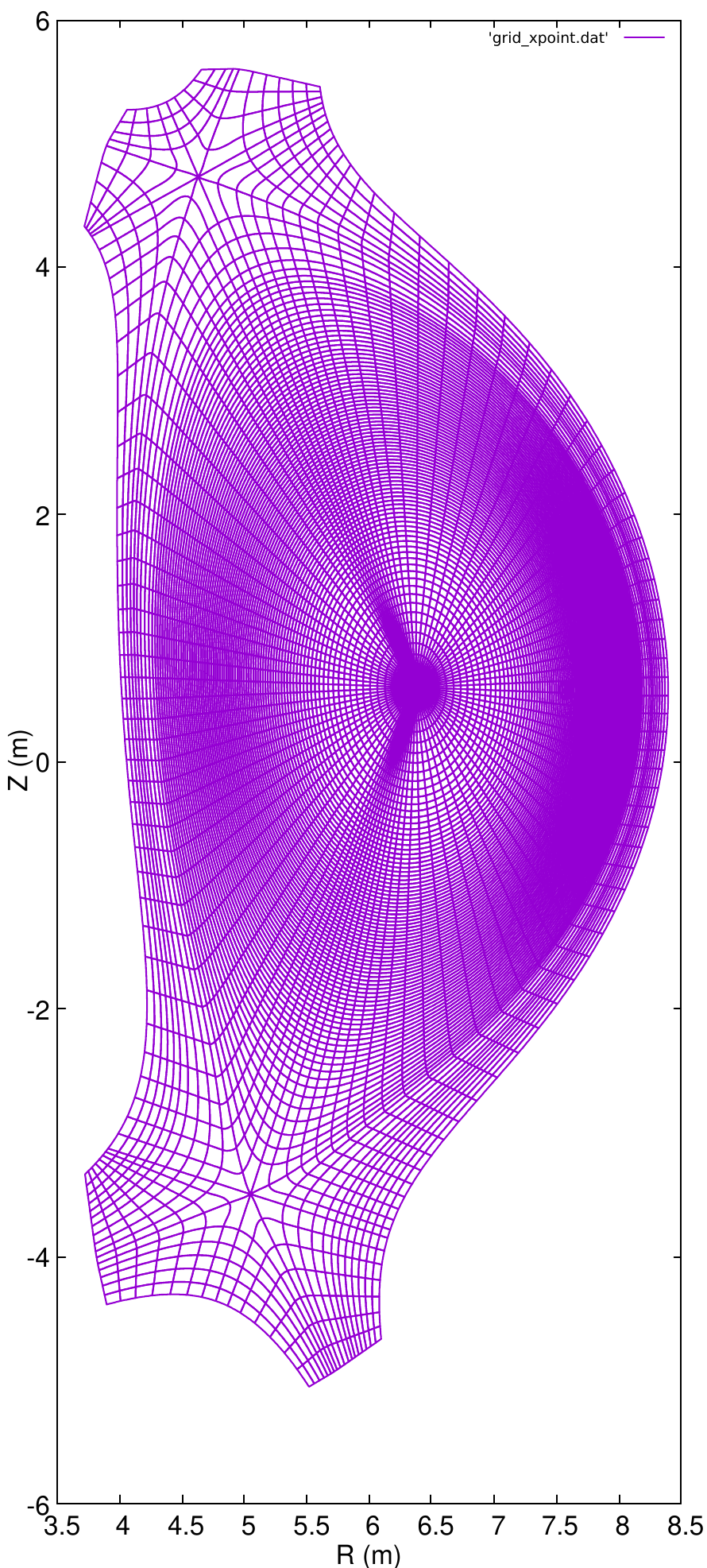}
}
\etbl
\caption{The simulation grid. The boundary is assumed to be ideally conducting.}
\label{fig:GridXpoint}
\end{figure*}

The grid we use is presented in Fig.\,\ref{fig:GridXpoint}, where the boundary approximately represents the position of the thin wall equivalent of the ITER blanket modules \cite{Villone2010NF}. We assume the boundary to be ideally conducting in this study. We include $12$ toroidal harmonics from $n=0$ to $n=11$ in our simulations, and our ablative density source is artificially elongated along the toroidal direction due to limitation in the toroidal resolution. This artificial toroidal spread with $\gD\gj_{NG}=0.3$ radian, corresponding to $2.46m$ to $1.86m$ as the fragments travel from the edge to the core, enable us to keep the Fourier components up to the first order in strength with the following density deposition shape:
\bqbl
\lbq{eq:GaussianSource}
S_n
\propto
\exp{\left(-\R{\left(R-R_{f}\right)^2+\left(Z-Z_{f}\right)^2}{\gD r^{2}_{NG}}\right)}
\times\exp{\left(-\left(\R{\gj-\gj_{f}}{\gD\gj_{NG}}\right)^2\right)}
.\eqbl
Here $R_f$, $Z_f$ and $\gj_f$ are the coordinates of the ablating fragment. We choose $\gD r_{NG}=0.08m$ in this study, but wish to emphasize that the overlapping between the ablation clouds from different fragments tends to create an envelop of plasmoid, so that our result is insensitive to the exact choice of $\gD r_{NG}$ so long as it is no larger than the plume width. The exact value of the density source is determined by the Neutral Gas Shielding (NGS) ablation model \cite{Zhang2020NF}, the detailed implementation is described in Ref. \cite{Di2023NF}. The density generated by ablation would then relax by parallel convection driven by the pressure gradient along the field lines, following the continuity equation.

The two temperature model described in Ref. \cite{Di2023NF} is used, where we track the evolution of the electron and the ion temperatures separately. The Braginskii parallel heat conduction \cite{Braginskii1965RPP} is used for each species respectively. We impose an upper limit of this parallel heat conduction corresponding to the value of the Braginskii electron thermal conductivity at $T_e\simeq 1.6keV$, beyond which the heat conductivity would not rise any more. This is an attempt to mimic the flux limit of the free streaming electrons and to prevent the overestimation of Braginskii thermal conduction in the high temperature regime. With the initial density, the perpendicular particle diffusion is set to be constant at $6m^2/s$ and the perpendicular heat conduction at $10m^2/s$. The choice of these coefficients are meant approach the gyro-Bohm scaling \cite{ITERProgress} while still maintain numerical stability. We do not expect the exact value to impact our result drastically since the stochastic transport and the macroscopic convective transport dominate over the turbulent one after the injection for our simulations. The perpendicular viscosity is set to $30m^2/s$ and the parallel one to $150m^2/s$, these values are chosen to meet the requirement of numerical instability suppression. The Spitzer resistivity \cite{SpitzerBook} with the effective charge contribution \cite{Hirshman1978POF} included is used for our simulation. Here, we have neglected the neo-classical resistivity contribution from the trapped electrons as the temperature drop and the density rise after the injection would prevent the electrons to complete a bounce period before collision. This is consistent with the previous finding that the mean-free-path of several hundred $eV$ electrons is comparable with the size of the cold, dense plasmoid around the fragments after ITER SPIs \cite{Di2023NF}. We assume the plasma is transparent to the impurity radiation, hence our result might over-estimate the peak radiation heat flux due to neglecting the plasma opacity.

The SPIs are carried out from the outer equatorial ports EQ-08 and EQ-17, which locate at $R=8.4m$, $Z=0.685m$ and opposite toroidal positions, along the major radial direction. The vertex angle of the injection cone is $20$ degrees. The fragment size distribution follows that of the Statistical Fragmentation model \cite{ParkDistribution,Gebhart2020FST}. Our two sets of SPI parameters are the ``$1\%$ neon full pellet SPI'' dual-SPIs with $2.5\times10^{22}$ neon atoms and $1.8\times10^{24}$ hydrogen atoms for each injector, as well as the ``$5\%$ neon quarter pellet SPI'' dual-SPIs with $2.5\times10^{22}$ neon atoms and $4.5\times10^{23}$ hydrogen atoms for each injector. For the full pellet case, the pellet is shattered into 300 fragments, while for the quarter pellet case the fragment number is 100. Their characteristic fragment size is $\gk_p^{-1}\simeq1.43mm$ and $\gk_p^{-1}\simeq1.30mm$ respectively. Both cases use $500m/s$ reference injection velocity with $\pm 40\%$ velocity spread. We assume the fragment velocity distributes uniformly over the spreading range, although in reality the velocity distribution has a correlation with the fragment size \cite{Gebhart2021FST}. We do not expect this to strongly alter the dynamics we investigate here, since one could usually modify the shattering angle to partly cancel this effect. It should also be noted that, for our baseline H-mode case, we enabled the velocity spread only in the EQ-08 plume and set all fragments in the EQ-17 plume with the same velocity value, albeit with correct direction spread so that there exists an asymmetry in the dual-SPI despite the two plumes being injected at the same time. We nonetheless don't expect this to greatly impact our results, since in reality there would always exist some asymmetry between the plumes even with very good synchronization between the injectors. 

\begin{table*}
\centering
\small
\noindent
\btbl{|c|c|c|c|c|c|}
\hline
Notation & Equilibria & Neon & Hydrogen & Frag. &  Delay\\
\hline
BH-FP-dt0 (case 1) & baseline & $2\times2.5\times 10^{22}$ & $2\times1.8\times 10^{24}$ & $300$ & $0ms$ (Asymm.)\\
\hline
DH-FP-dt0 (case 2) & degraded & $2\times2.5\times 10^{22}$ & $2\times1.8\times 10^{24}$ & $300$ & $0ms$\\
\hline
DH-QP-dt0 (case 3) & degraded & $2\times2.5\times 10^{22}$ & $2\times4.5\times 10^{23}$ & $100$ & $0ms$\\
\hline
DH-QP-dt1 (case 4) & degraded & $2\times2.5\times 10^{22}$ & $2\times4.5\times 10^{23}$ & $100$ & $1ms$\\
\hline
\etbl
\caption{The injection parameters for the SPI considered in this study. Note that for BH-FP-dt0 there exists an asymmetry between the plumes although they are injected at the same time.}
\label{tab:1}
\end{table*}

We have summarized the cases we investigate in this paper in Table \ref{tab:1}, where the target equilibria, the neon and hydrogen injection amount, the fragment number and the delay between the injectors are shown. For all cases, the first fragment of the EQ-08 plume arrives on the LCFS approximately at $t=0.38ms$ and the last one arrives approximately at $0.89ms$. All the fragments from the EQ-17 plume in the BH-FP-dt0 case arrive approximately at $t=0.54ms$ since there is no velocity spread. The arrival time of the fragments from the EQ-17 plume for the rest of the cases is the same with that of the EQ-08 plume, plus the delay between the injectors shown in Table \ref{tab:1}. We will mainly investigate BH-FP-dt0, DH-FP-dt0 \& DH-QP-dt0 in Section \ref{s:PlasmoidDrift}, and BH-FP-dt0, DH-QP-dt0 \& DH-QP-dt1 in Section \ref{s:HeatFlux}.

\subsection{The energy impact}
\label{ss:EnergyImpact}

To estimate the wall material temperature rise in an inertially cooled scenario, it is important to calculate the accumulated heat on the wall surface considering both the incoming heat flux and the conductive cooling of the PFCs. For the simple case of 1D heat conduction concerning a semi-infinite slab and constant heat flux $q_s$, the temperature increase $\gD T$ after $\gD t$ time of heat exposure is \cite{Herrmann2002PPCF}
\bqbl
\lbq{eq:TIncrease}
\gD T(\gD t)
=
\R{2}{\sqrt{\gp}b}\gD Q (\gD t)
=
\R{2}{\sqrt{\gp}b}q_s \sqrt{\gD t}
.\eqbl
Here $b=\sqrt{\gk \gr c}$, $\gr$ is the material density, $c$ is its specific heat capacity, $\gk$ is its heat conductivity. The temperature rise is proportional, via the material-dependent factor $b$, to the so-called energy impact $\gD Q\equiv q_s\sqrt{\gD t}$ which itself is material-independent. The maximum tolerable energy impact could then be obtained for each individual material \cite{Herrmann2002PPCF,Pitts2015JNM}, thus providing insights into the probability of severe material damage for a given period of heat deposition.

When considering a time varying heat flux which is more relevant during transient TQ process, one could use the fundamental solution of the heat equation to obtain an analytical description of the heat accumulation on the slab surface without numerically solving the 1D heat conduction equation. This is equivalent to considering the heat pulse $q_s(t)$ as a series of continuously arriving ``heat packets'', each of which individually undergoes a 1D random walk towards infinity with step-length determined by the material heat conductivity after their arrival. The total heat distribution after a certain time period $\gD t$ is then the summation of the 1D spatial probability distribution of all the packets. For each such heat packet, the fundamental solution on $x\in\left[0,+\infty\right)$ is
\bqbl
\lbq{eq:FundamentalSolution}
\gF\left(x,\gD t\right)
=
\R{1}{\sqrt{\gp \ga \gD t}}\exp{\left(-\R{x^2}{4\ga \gD t}\right)}
.\eqbl
Here $\ga$ is the thermal diffusivity and $x$ is the distance along the 1D direction. Since we are only concerned with the temperature rise at the wall surface, $x$ could be practically set to zero, and we obtain the time decay of each individual heat packet to be proportional to $\gD t^{-1/2}$. The energy impact integrating over all the heat packets could then be obtained by the following convolution:
\bqbl
\lbq{eq:EnergyImpact}
\gD Q (t)
=
\R{1}{2}\int_{t_0}^{t}{\R{q_s(t')}{\sqrt{t-t'}}dt'}
.\eqbl
Here $t_0$ is the beginning time of the heat pulse. For a constant $q_s$, one naturally recovers the previous result $\gD Q (\gD t) = q_s\sqrt{\gD t}$. Henceforth we will calculate the energy impact using Eq.\,(\rfq{eq:EnergyImpact}) and estimate the potential heat damage based on it.

\section{Plasmoid drift, accompanied heat flux and its mitigation}
\label{s:PlasmoidDrift}

\begin{figure*}
\centering
\noindent
\btbl{ccc}
\parbox{1.65in}{
	\includegraphics[scale=0.24]{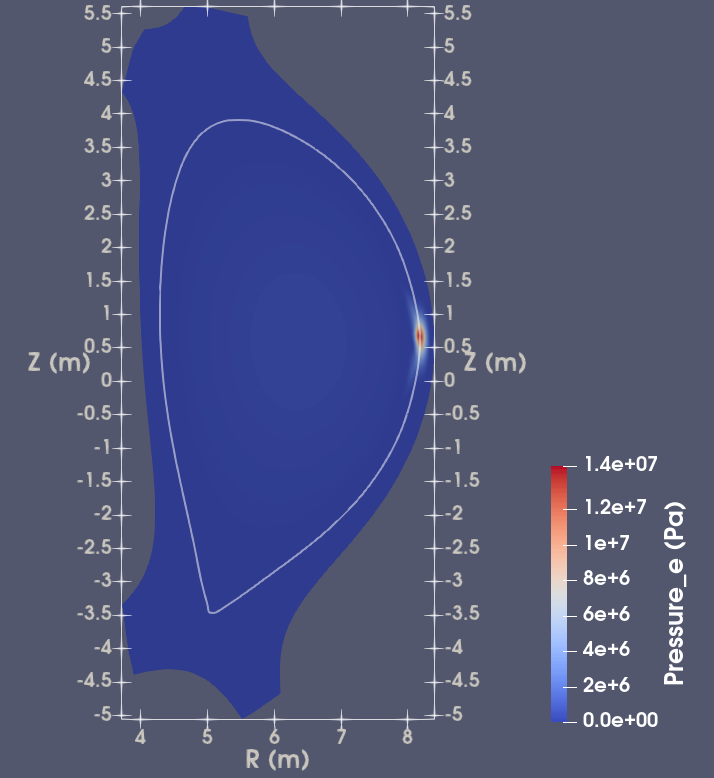}
}
&
\parbox{1.65in}{
	\includegraphics[scale=0.24]{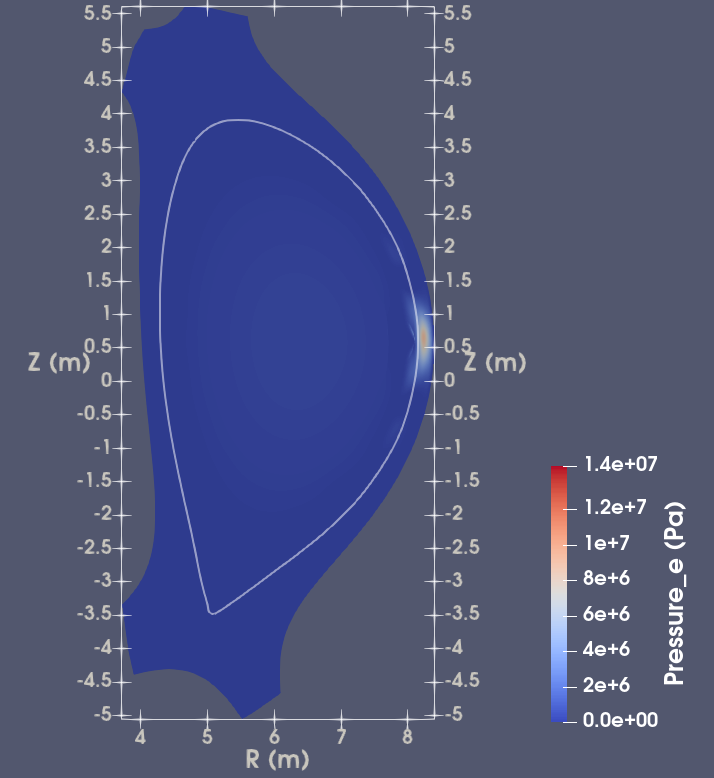}
}
&
\parbox{2.5in}{
	\includegraphics[scale=0.24]{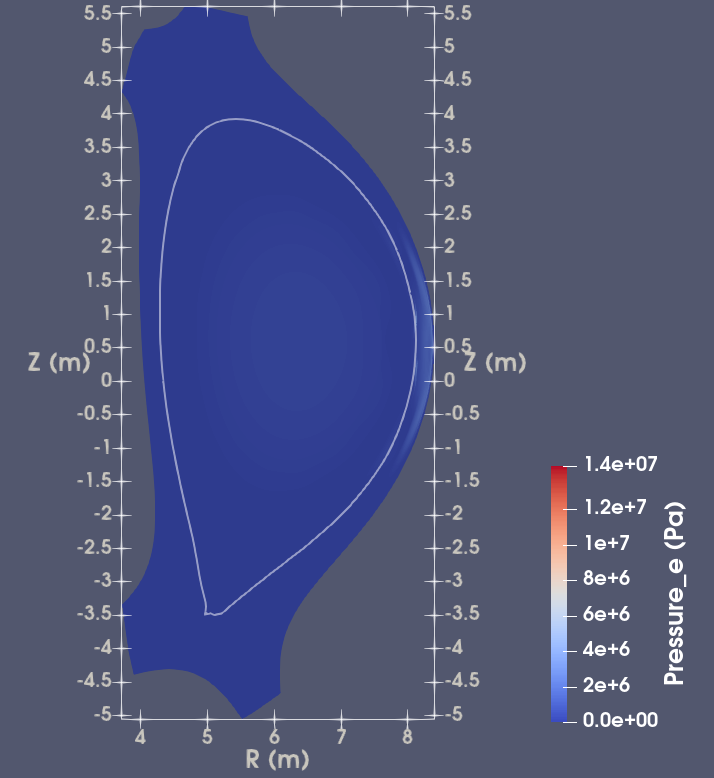}
}
\\
(a)&(b)&(c)
\etbl
\caption{The electron pressure profile at the same toroidal location as the EQ-08 plume for BH-FP-dt0 at (a) $t=0.41ms$, (b) $t=0.42ms$ and (c) $t=0.46ms$. The white contour indicates the position of the equilibrium LCFS. The over-pressure as a result of the vanguard fragments arriving on the pedestal as well as its fast outward drift and expulsion can be identified.}
\label{fig:PlasmoidDrift}
\end{figure*}

For the baseline H-mode BH-FP-dt0 studied here, it is found that the high thermal energy reservoir could result in strong local pressure peak around the vanguard fragments, which induces vertical polarization, which in turn results in $\waE\times\waB$ drift along the major radial direction \cite{Rozhansky1995PPCF}. 
In our simulations, the polarization could be implicitly obtained through the evolution of the vorticity equation in Ref. \cite{Di2023NF}: 
\bqbl
\lbq{eq:VorticityEq}
R\na\cdot\left[R^2\R{\pa}{\pa t}\left(\gr\na_{pol}u\right)\right]
&
=
&
\R{1}{2}\left\{R^2\left|\na_{pol}u\right|^2,R^2\gr\right\}
+\left\{R^4\gr\gw,u\right\}
\nonumber
\\
&&
-R\na\cdot\left[R^2\na_{pol}u\na\cdot\left(\gr\wsv\right)\right]
+\left\{\gy,j\right\}
\nonumber
\\
&
&
-\R{F_0}{R}\R{\pa j}{\pa\gj}
+\left\{P,R^2\right\}
+R^3\gm_\bot\left(T_e\right)\na_{pol}^2\gw
.\eqbl
Here $\gr$ is the mass density, $u$ is the stream function $\wsv=v_\|\waB-R^2\na u\times\na \gj$. To see this, one realizes that the stream function $u$ is associated with the electric potential, and the second term on the RHS of the velocity definition is essentially the $\waE\times\waB$ drift. Thus the Poisson bracket $\left\{P,R^2\right\}$ term in Eq.,(\rfq{eq:VorticityEq}), which includes the $\na B$ drift contribution, would drive the aforementioned polarization in the presence of the local plasmoid over-pressure. Such polarization would then in turn result in the drift motion.
The above discussed strong drift motion is found when the fragments arrive on the pedestal of the baseline H-mode. It should be noted that due to the asymmetry in our two plumes as mentioned in Section \ref{ss:System}, the EQ-08 plume arrives on the pedestal first and triggers the strong drift. The resulting drift motion expels significant amounts of injected materials out of the Last Closed Flux Surface (LCFS), undermining the injection assimilation, as can be seen in Fig.\,\ref{fig:PlasmoidDrift}, which shows to the same toroidal location as the EQ-08 plume. The local electron over-pressure could be readily identified at $t=0.41ms$ in Fig.\,\ref{fig:PlasmoidDrift}(a) as the fragments penetrate the equilibrium LCFS represented by the white contour. This over-pressure region essentially is the envelop of several plasmoid produced by individual vanguard fragments as they enter the pedestal. This high pressure plasmoid envelop is then seen convected outward in Fig.\,\ref{fig:PlasmoidDrift}(b) due to the major radial drift and finally relaxes outside of the separatrix in Fig.\,\ref{fig:PlasmoidDrift}(c). 

\begin{figure*}
\centering
\noindent
\btbl{cc}
\parbox{2.45in}{
	\includegraphics[scale=0.15]{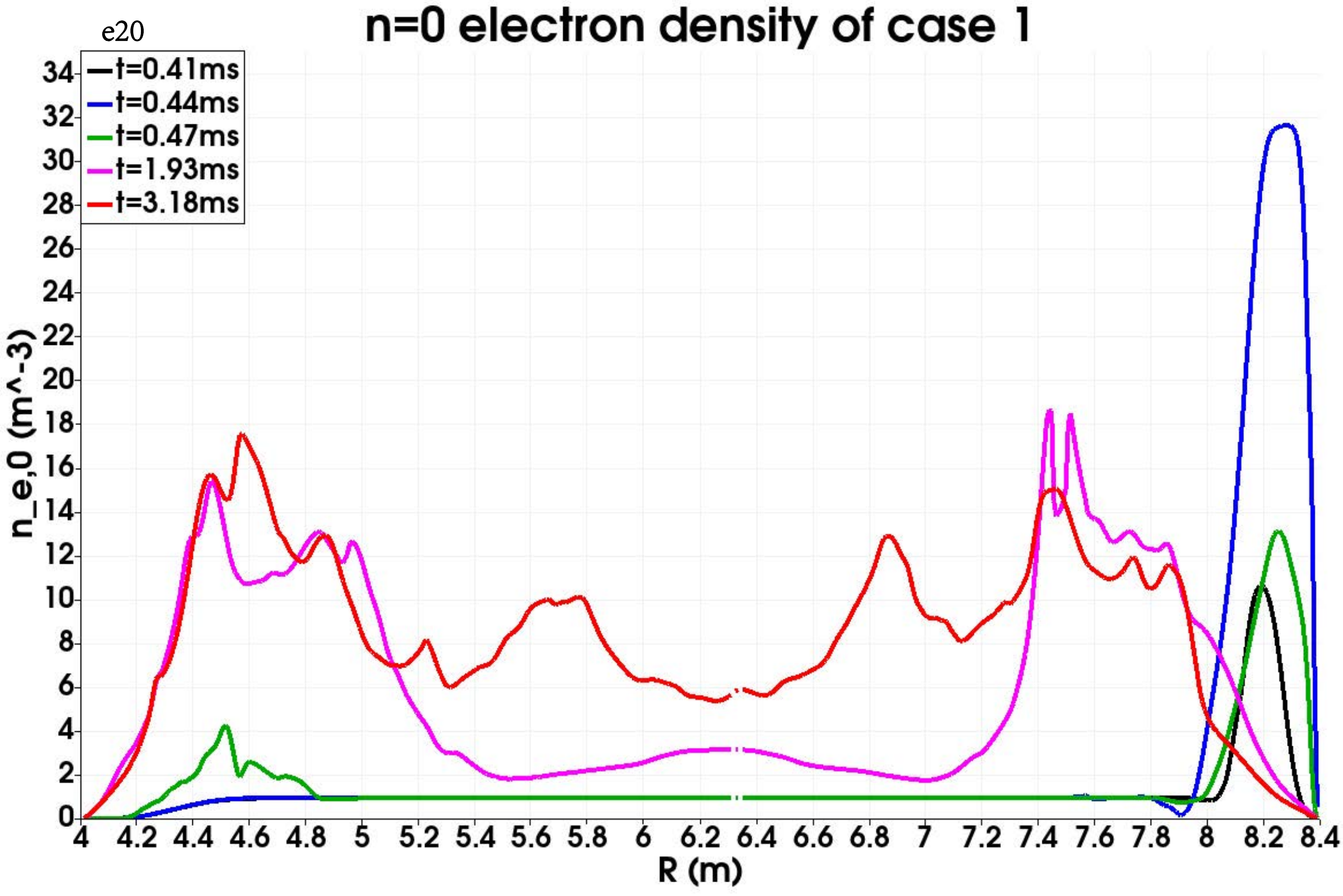}
}
&
\parbox{2.45in}{
	\includegraphics[scale=0.15]{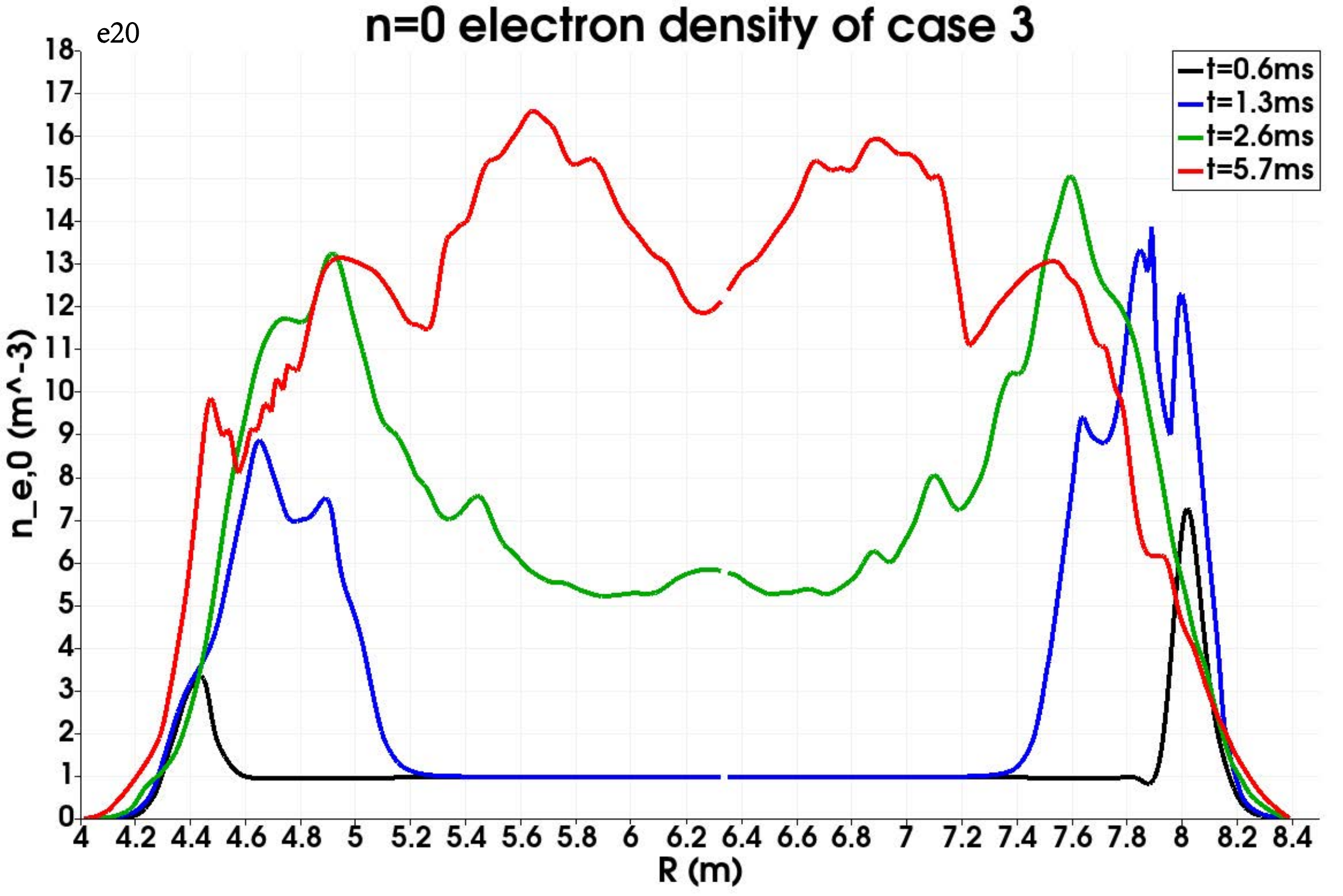}
}
\\
(a)&(b)
\etbl
\caption{The mid-plane cut of the $n=0$ electron density for (a) BH-FP-dt0 and (b) DH-QP-dt0 at various times. The high pedestal temperature and the consequential plasmoid drift in BH-FP-dt0 results in significant density deposition outside of the pedestal in the early injection phase as is shown by the black, blue and green lines in (a). The LCFS on the low field side approximately locates at $R=8.2m$.}
\label{fig:ne_profile}
\end{figure*}

Consequentially, the injection assimilation within the Last Close Flux Surface (LCFS) suffers in the early injection phase as is shown in Fig.\,\ref{fig:ne_profile}, where the comparison between the mid-plane cut of the $n=0$ electron density profile for BH-FP-dt0 and DH-QP-dt0 at various times is shown. The LCFS on the low field side approximately locates at $R=8.2m$ for both cases. In Fig.\,\ref{fig:ne_profile}(a), comparing the black, blue and green lines, it can be seen that the density peak initially occurs near the LCFS, then goes outward into the scrape-off layer. This drift motion, coupled with the associated outgoing heat flux, causes significant density deposition in the open field line region, which is then lost over time. As time goes on however, the fragments ultimately penetrate deeper into the plasma and result in core density rise in the late injection phase, as shown by the cyan and red lines. Note that the density rise in the axis region is due to inward MHD transport, as at that time the majority of fragments has not reached the axis yet. This behaviour is compared with DH-QP-dt0 shown in Fig.\,\ref{fig:ne_profile}(b), where no apparent density expulsion is seen and the density peak gradually moves inward.

\begin{figure*}
\centering
\noindent
\btbl{cc}
\parbox{2.45in}{
	\includegraphics[scale=0.45]{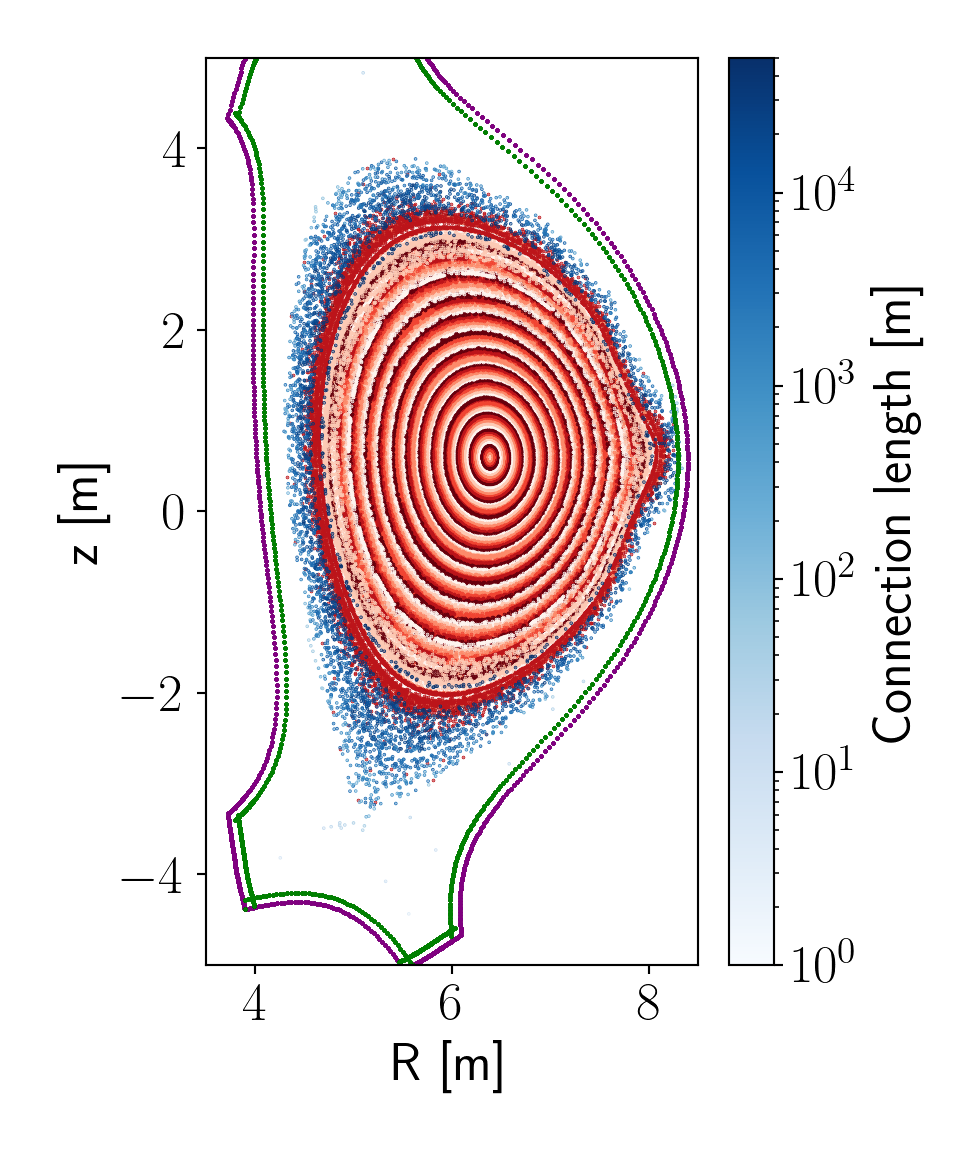}
}
&
\parbox{2.45in}{
	\includegraphics[scale=0.45]{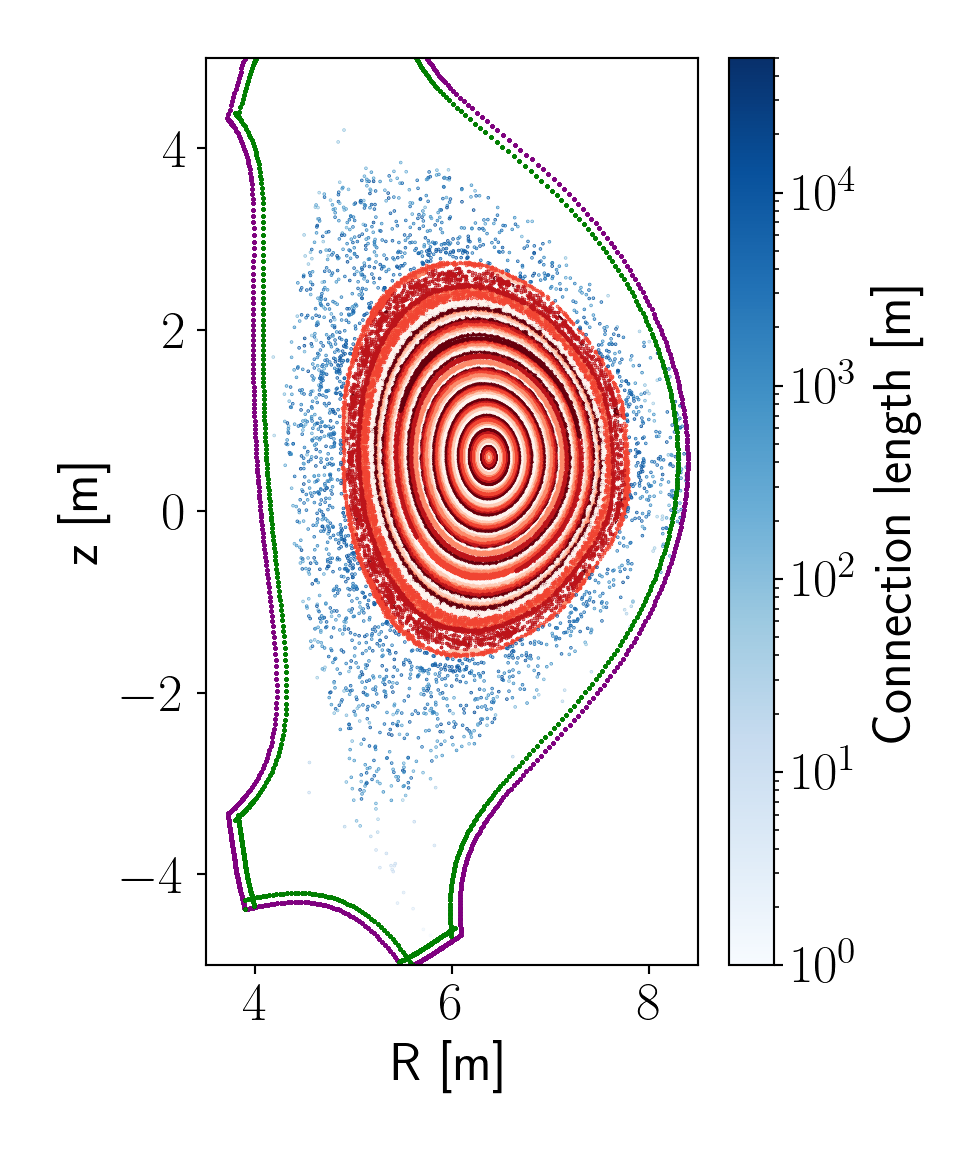}
}
\\
(a)&(b)
\etbl
\caption{Poincar\'{e} plots of the magnetic field lines at time (a) $t=0.41ms$ and (b) $t=0.46ms$ for BH-FP-dt0. The field lines remain well confined after $2000$ toroidal turns are marked by red, while those that are lost are coloured according to their ``loss length''. The outward deformation of the flux surfaces on the mid-plane in (a) is likely due to the outward plasma motion.}
\label{fig:DriftPoincare}
\end{figure*}

Such drift motion is accompanied by strong field line stochasticity in the edge region, resulting in a stochastic parallel heat flux, as well as perpendicular convective heat flux due to the drift transport of the high pressure plasmoid. Poincar\'{e} plots of the magnetic field lines at $t=0.41ms$ and $t=0.46ms$ for BH-FP-dt0, as well as the simulation boundary and a surface $10cm$ inside of that boundary which corresponds to the approximate position of the first wall, are shown in Fig.\,\ref{fig:DriftPoincare}, where the field lines that remain well confined after $2000$ toroidal turns are marked by red, while those that are lost are coloured according to the distance they travel before they hit the simulation boundary. 
The figure corresponds to the same toroidal location as the EQ-08 plume.
The simulation boundary is shown in purple, and the first wall approximation is shown in green. In Fig.\,\ref{fig:DriftPoincare}(a), the edge region begins to become stochastic, and the ``connection length'' of the field lines initialized in the edge region to the wall decreases. Another notable feature is the outward deformation of the closed flux surfaces close to the plasmoid position, likely a result of the outward plasma motion. In Fig.\,\ref{fig:DriftPoincare}(b), already significant region around the previous pedestal has become stochastic with short connection length to the wall, and the H-mode pedestal is collapsed. Comparing Fig.\,\ref{fig:ne_profile}(a) and Fig.\,\ref{fig:DriftPoincare}(b), it can be seen at $t=0.46ms$ there is no significant density assimilation within the pedestal, while strong edge stochasticity and thus thermal energy loss already occurs.

\begin{figure*}
\centering
\noindent
\btbl{cc}
\parbox{2.4in}{
	\includegraphics[scale=0.25]{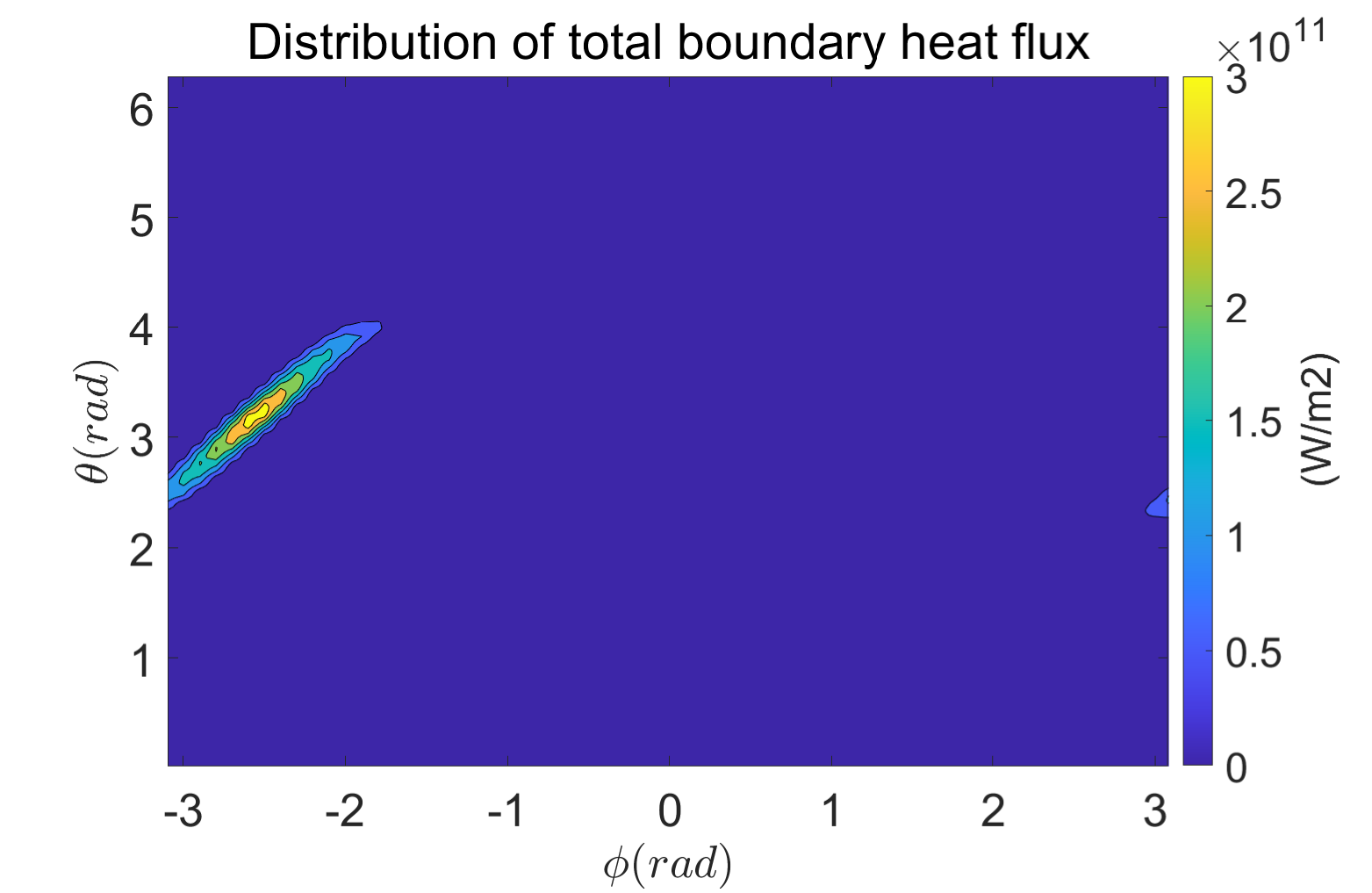}
}
&
\parbox{2.4in}{
	\includegraphics[scale=0.25]{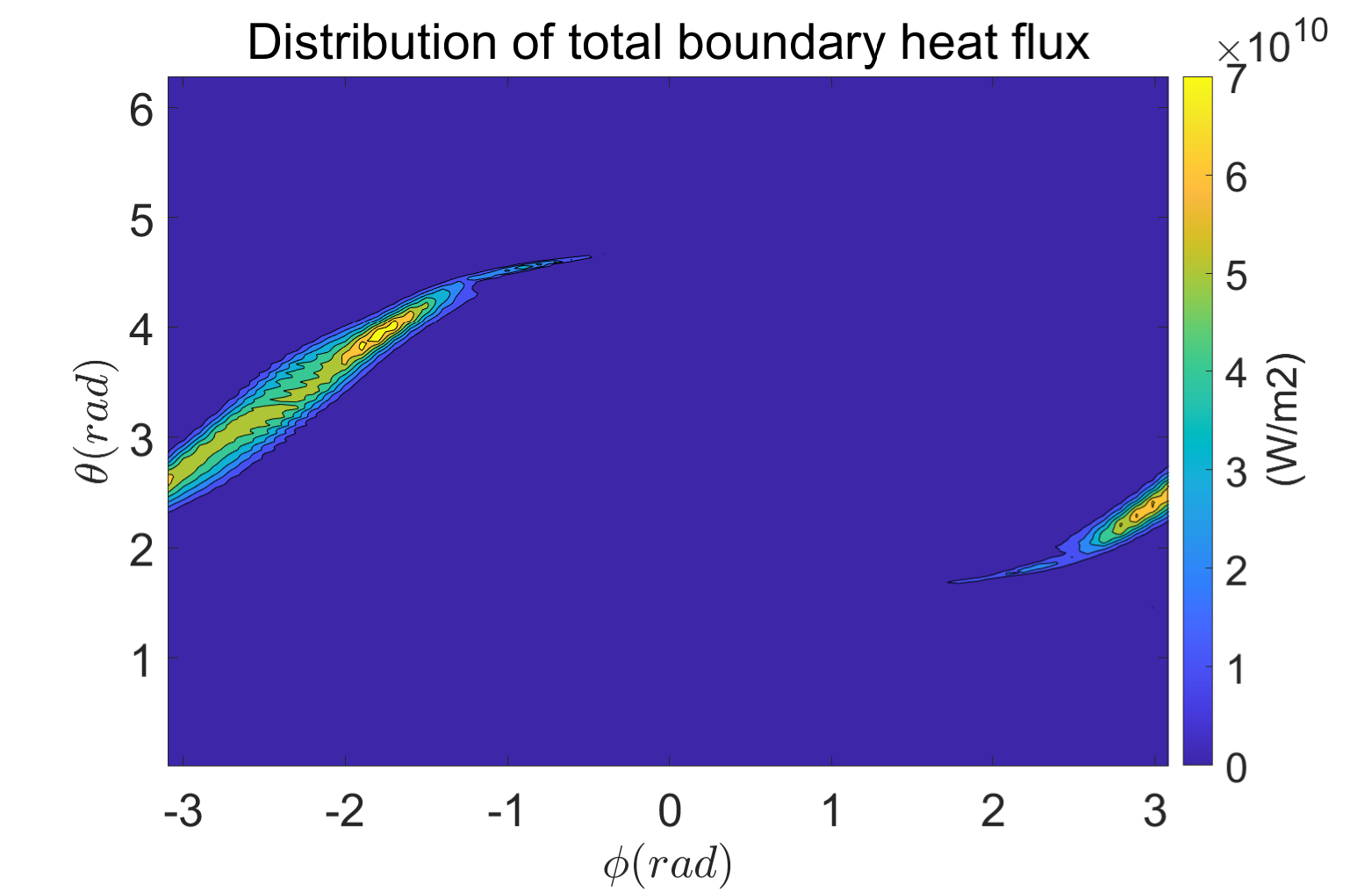}
}
\\
(a)&(b)
\etbl
\caption{The normal component of the combined conductive and convective heat flux for BH-FP-dt0 at time (a) $t=0.42ms$ and (b) $t=0.43ms$ across a surface $10cm$ within the simulation boundary, which represents the approximate location of the thin wall equivalent of the blanket module. The horizontal axis is the toroidal angle and the vertical one is the poloidal angle. The EQ-08 port locates approximately at $\gj=-2.57$. During this transient heat pulse the perpendicular heat flux dominates over the parallel one.}
\label{fig:FWConHeatFlux}
\end{figure*}

As a result of the above-mentioned stochastic and convective transport, the conductive and convective heat flux onto the first wall could be significant. In Fig.\,\ref{fig:FWConHeatFlux}, the normal component of the combined heat flux normal to the first wall approximation represented by the green line in Fig.\,\ref{fig:DriftPoincare} is shown for times $t=0.42ms$ and $t=0.43ms$. Comparing Fig.\,\ref{fig:PlasmoidDrift} and Fig.\,\ref{fig:FWConHeatFlux}, it can be seen that as the over-pressured plasmoid drifts towards the wall at $t=0.42ms$, a strong helical boundary heat flux structure begins to appear at the EQ-08 injection location centred around $\gj=-2.57$. It successively relaxes along field lines and the peak heat flux locations move away from the plasmoid drift location at $t=0.43ms$. During this transient heat pulse the perpendicular heat flux dominates over the parallel one.

\begin{figure*}
\centering
\noindent
\btbl{cc}
\parbox{2.4in}{
	\includegraphics[scale=0.21]{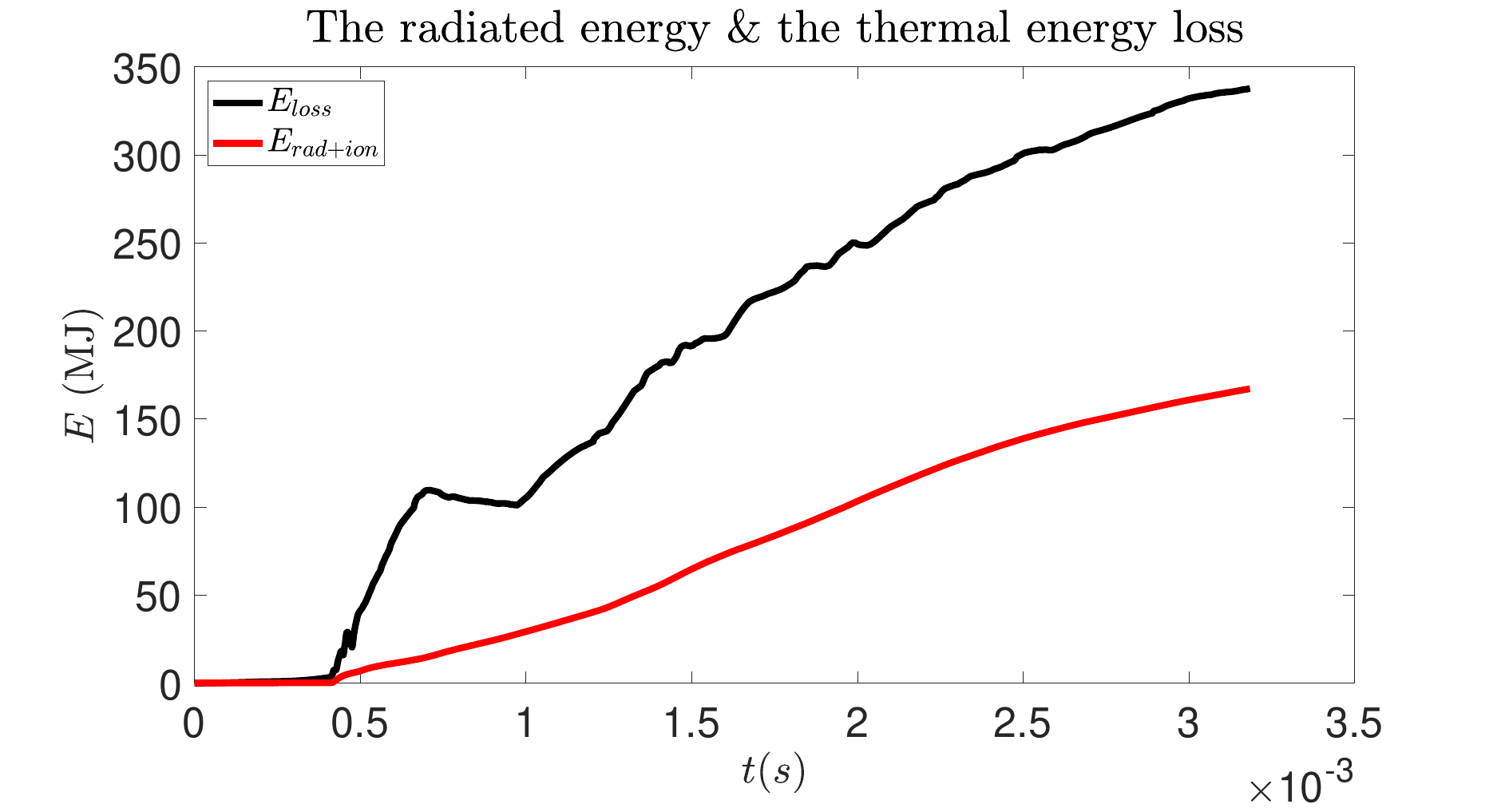}
}
&
\parbox{2.4in}{
	\includegraphics[scale=0.21]{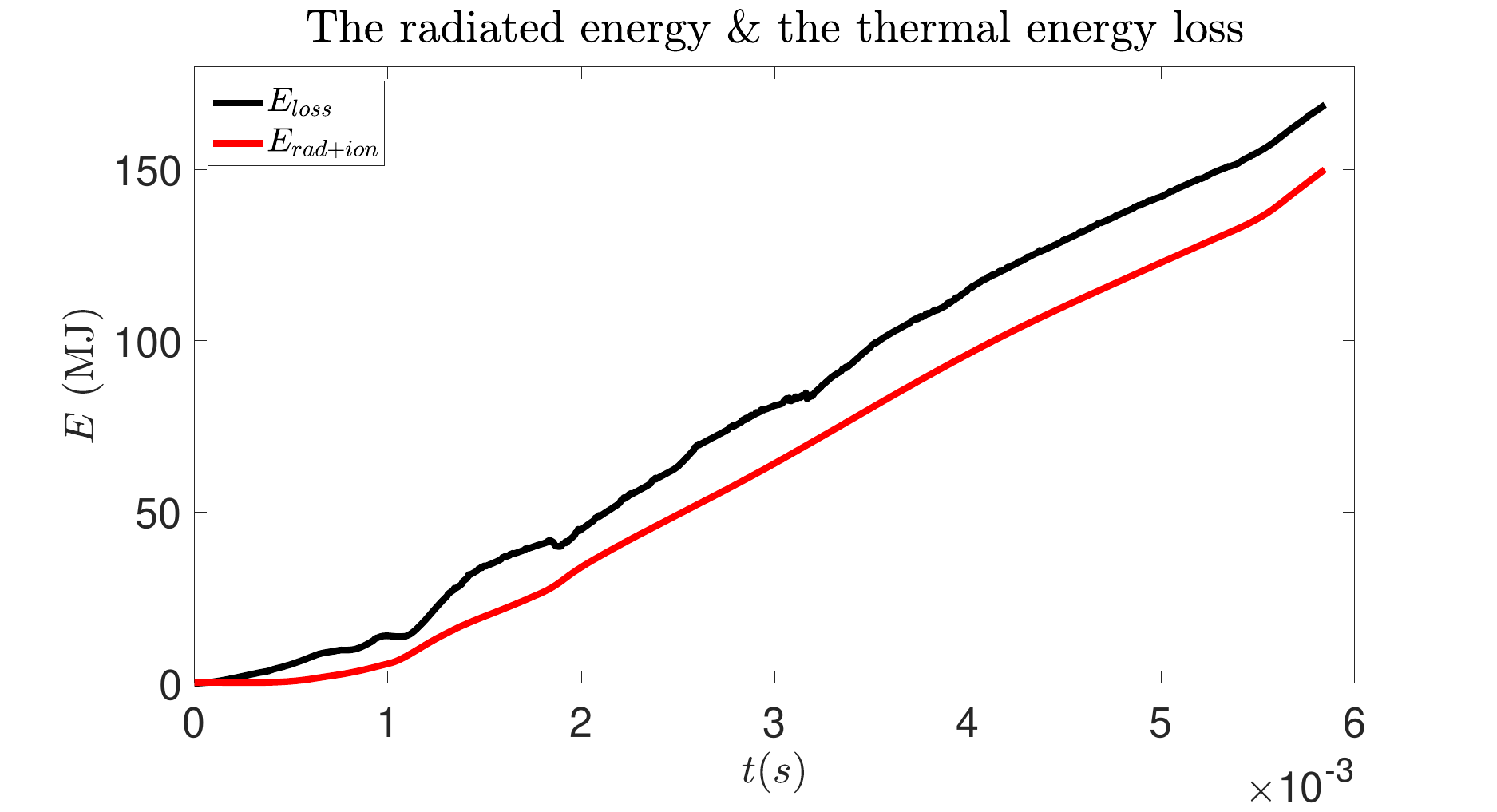}
}
\\
(a)&(b)
\etbl
\caption{Comparison of the total thermal energy loss with the energy consumed by radiation and ionization for (a) BH-FP-dt0 and (b) DH-QP-dt0.}
\label{fig:RadFraction}
\end{figure*}

Both the aforementioned density expulsion and the accompanying transport event are undesirable for TQ mitigation. The H-mode pedestal collapses after the transport event is triggered and over $100MJ$ of thermal energy is prematurely lost before the fragments penetrate deep into the core, compared with the $370MJ$ initial thermal energy. However, the plasma usually would already experience significant confinement degradation due to the growth of precursor modes before the disruption onset \cite{Riccardo2005NF} and the triggering of the SPIs. Hence it is interesting to see if the reduced thermal energy reservoir of such a ``degraded H-mode'' could still trigger such strong drift motion, and whether an increased neon mixture ratio could help to suppress the plasmoid drift by radiatively reducing the local over-pressure, and thus mitigate the accompanying outgoing heat flux. Indeed, if we compare the thermal energy loss and the radiated power of BH-FP-dt0 and DH-QP-dt0 as is shown in Fig.\,\ref{fig:RadFraction}, DH-QP-dt0 enjoys much better radiation fraction behaviour due to the mitigation of this drift motion. Here the total loss of the thermal energy is represented by the black solid lines, and the total radiated energy is represented by the red solid lines. For BH-FP-dt0, the sudden rise of the discrepancy between the thermal energy loss and the radiated energy beginning at $t=0.45ms$ is due to the aforementioned heat flux accompanying the drift motion. This heat flux is absent in DH-QP-dt0 as will be shown below. For both cases, the ohmic heating is negligible until the end of the simulation when it begins to pick up due to the thermal energy reservoir getting depleted.

\begin{figure*}
\centering
\noindent
\btbl{cc}
\parbox{2.4in}{
	\includegraphics[scale=0.25]{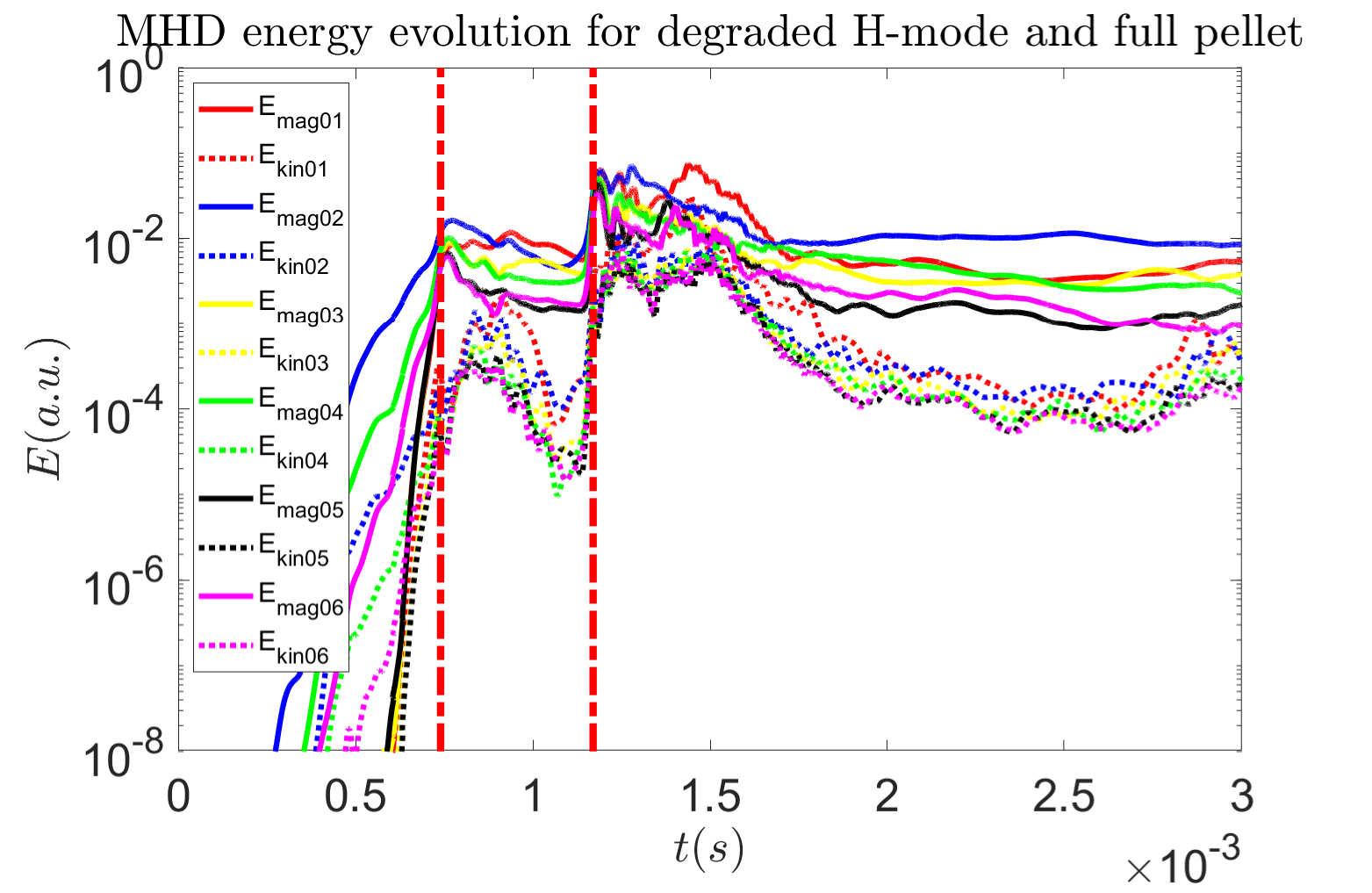}
}
&
\parbox{2.4in}{
	\includegraphics[scale=0.25]{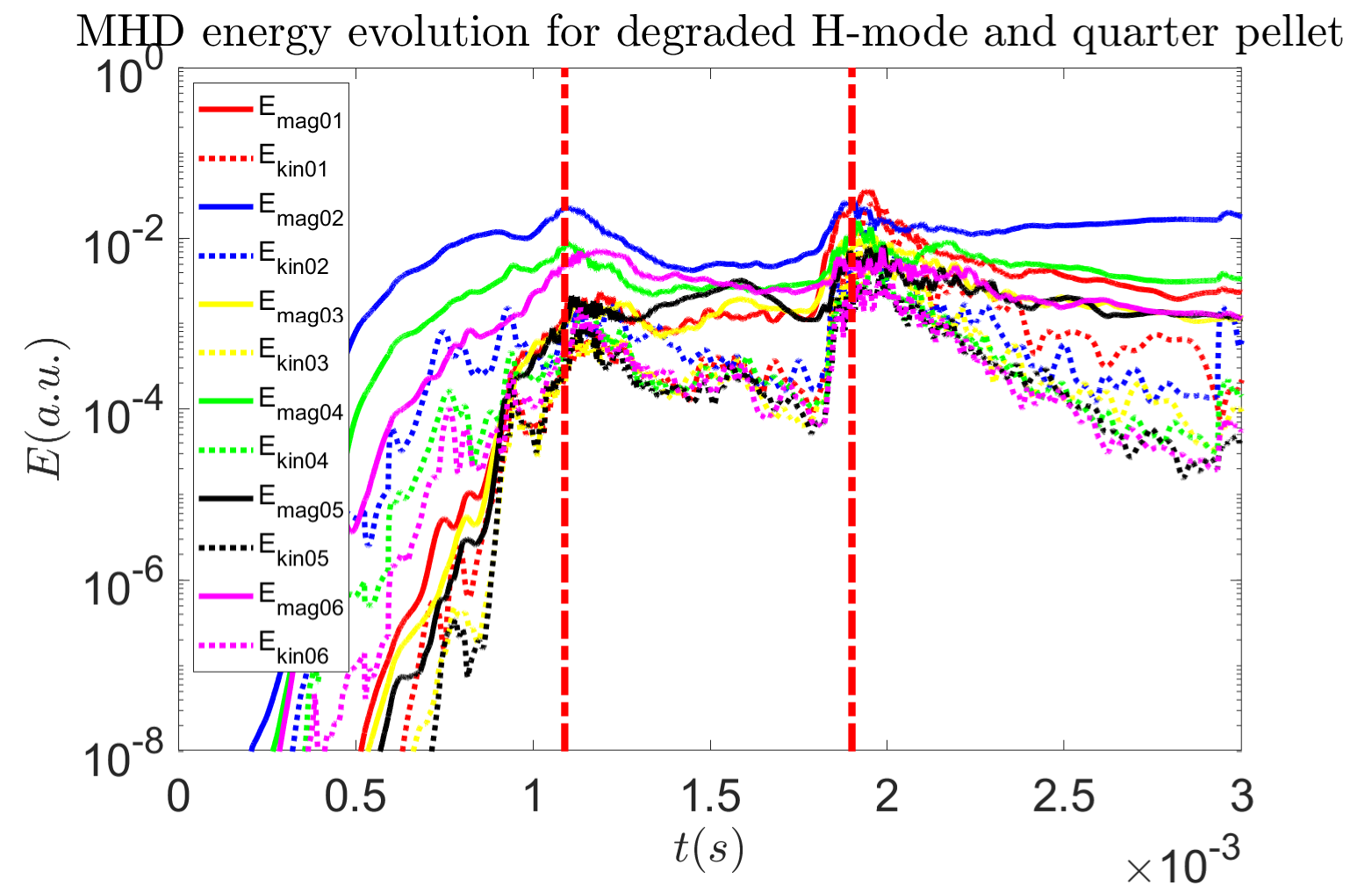}
}
\\
(a)&(b)
\etbl
\caption{The MHD spectrum of (a) DH-FP-dt0 and (b) DH-QP-dt0. The solid lines represent the perturbed magnetic energy and the dashed lines represent that of the kinetic energy. The red chained lines corresponds to the MHD peaks at the time of which we would investigate the boundary heat flux distribution.}
\label{fig:MHD_degraded}
\end{figure*}

The perturbed magnetic and kinetic energies with toroidal harmonics $n=1$ to $n=6$ for DH-FP-dt0 and DH-QP-dt0 are shown in Fig.\,\ref{fig:MHD_degraded}, where the perturbed magnetic energies are shown in solid lines and the kinetic ones are shown in dashed lines. We identified two MHD peaks in both cases, which are marked by the vertical chained lines in the figure. For both cases, the first peak corresponds to the fragment arrival on the plasma which results in edge stochastization, while the second peak corresponds to the core temperature relaxation. These strong MHD responses correspond to strongest outgoing heat flux and plasmoid drift motion, thus we would investigate the boundary heat flux at these times.

\begin{figure*}
\centering
\noindent
\btbl{cc}
\parbox{2.4in}{
	\includegraphics[scale=0.25]{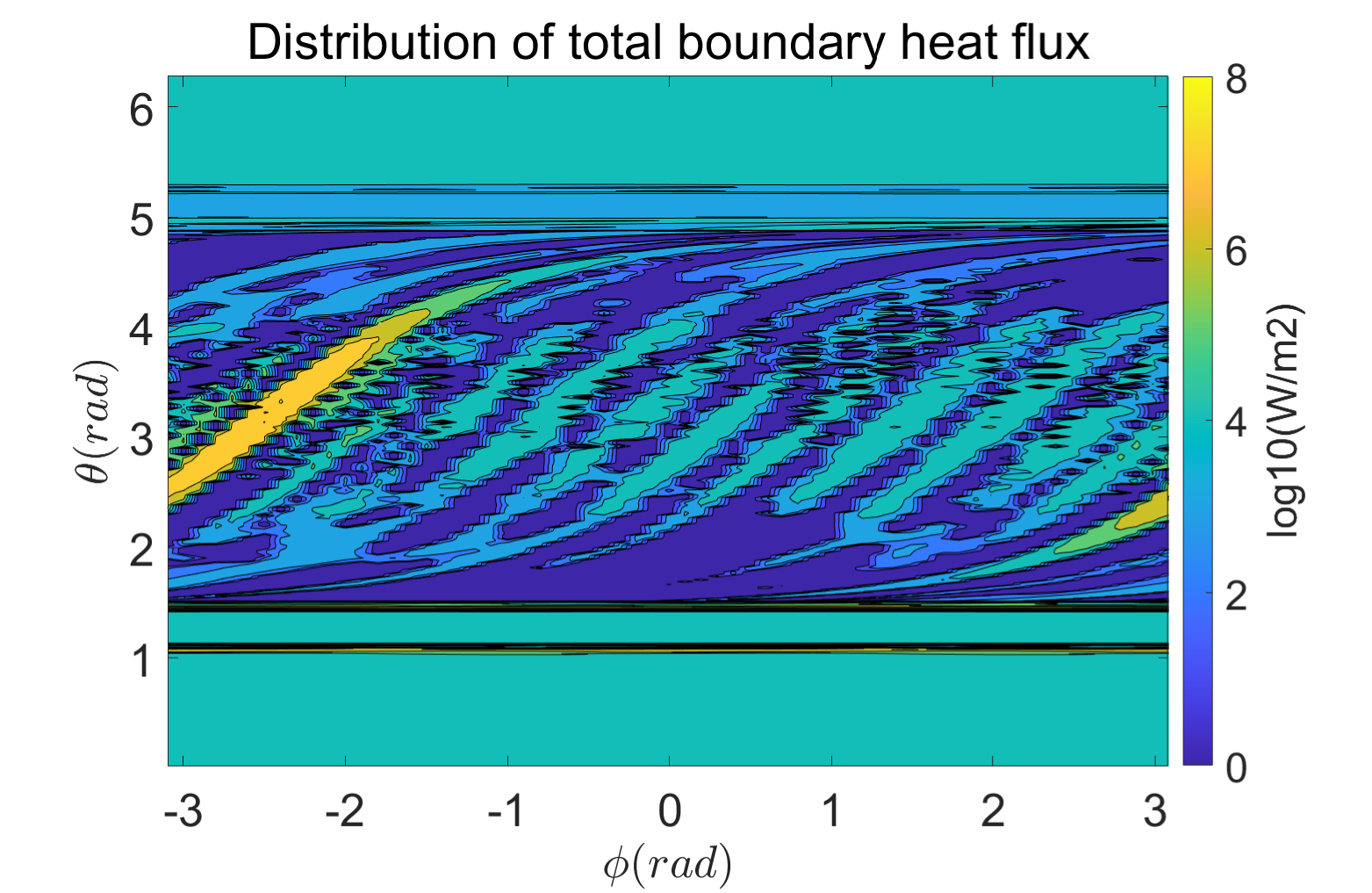}
}
&
\parbox{2.4in}{
	\includegraphics[scale=0.25]{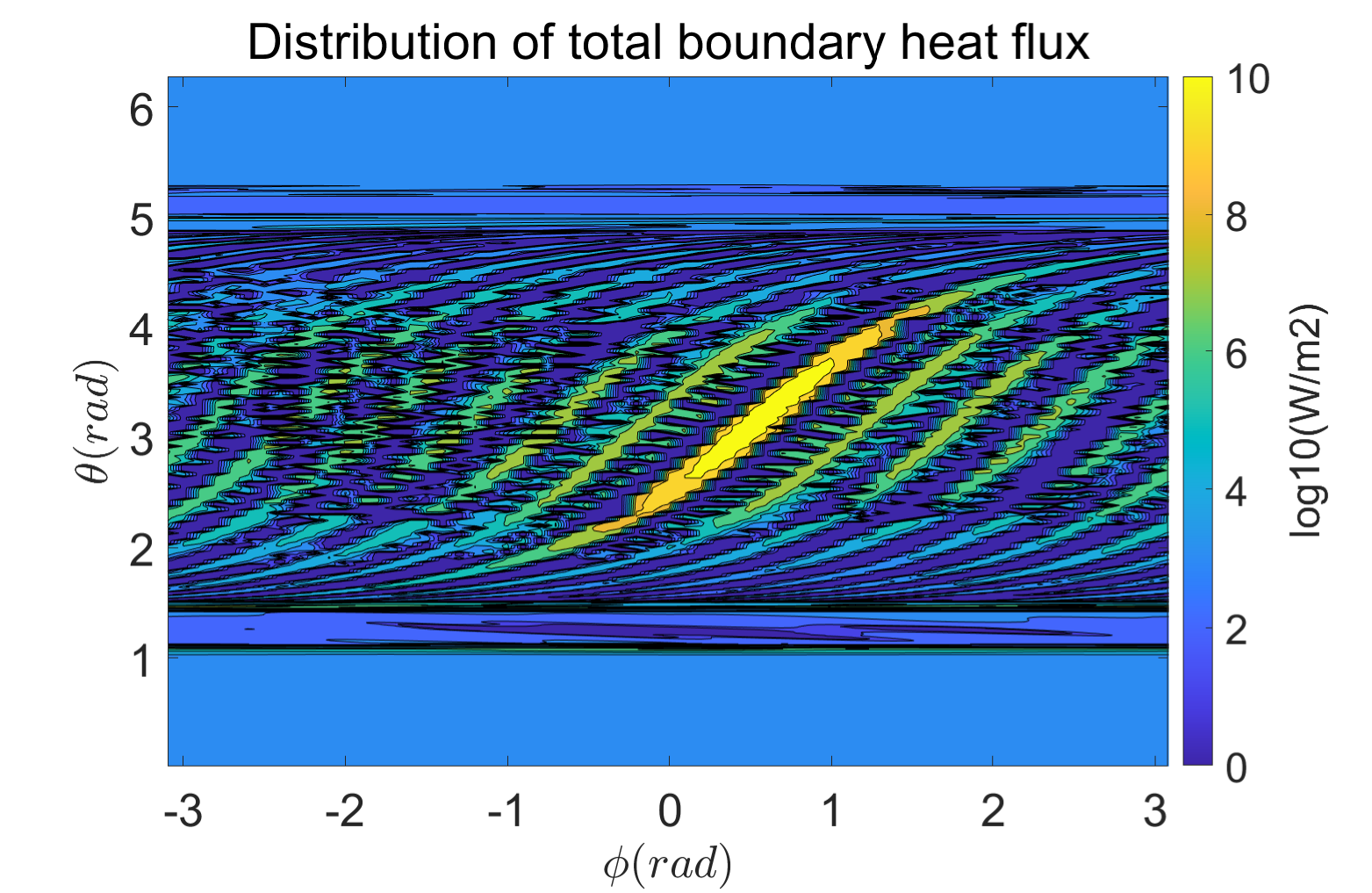}
}
\\
(a)&(b)
\\
\parbox{2.4in}{
	\includegraphics[scale=0.25]{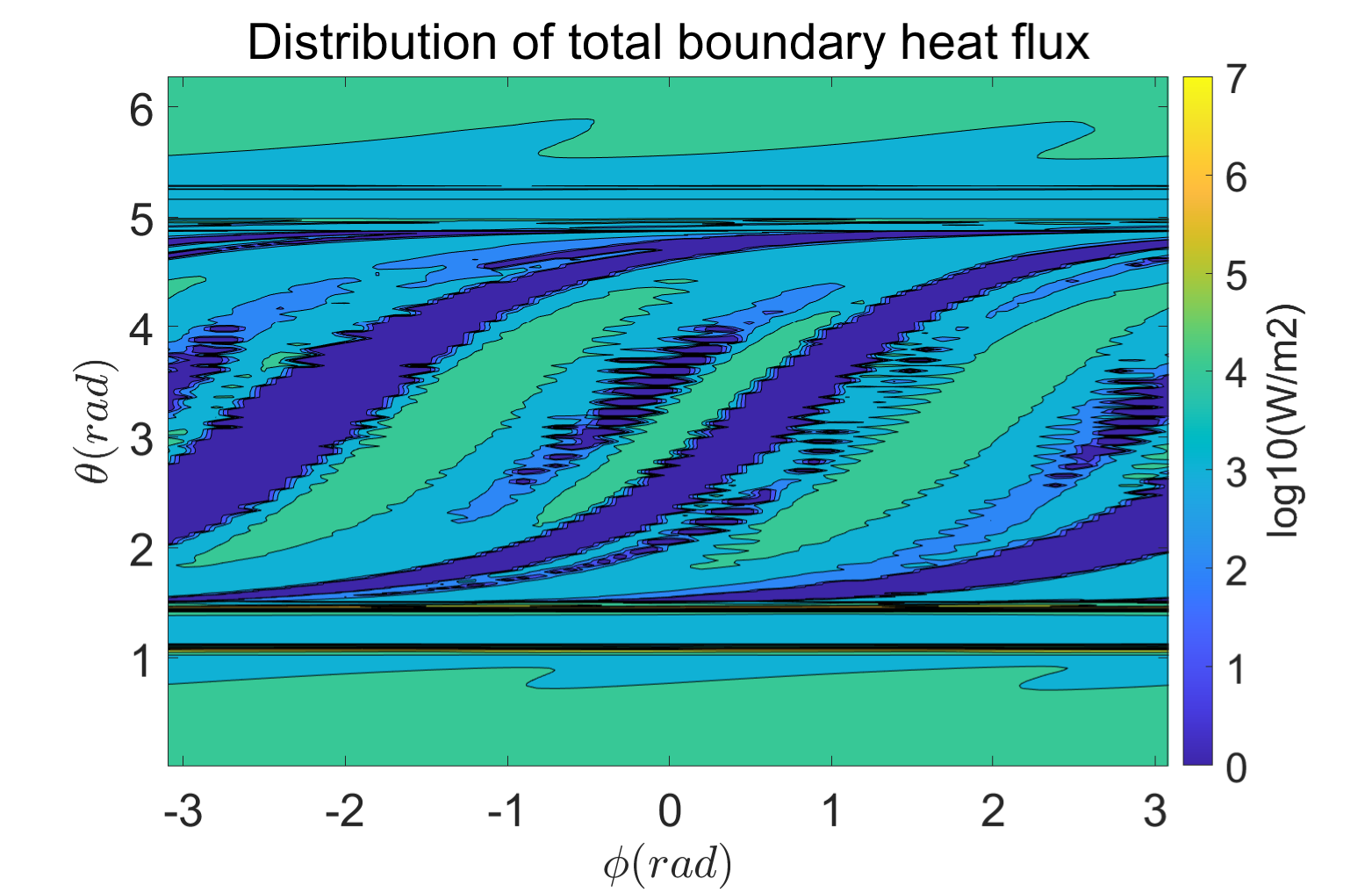}
}
&
\parbox{2.4in}{
	\includegraphics[scale=0.25]{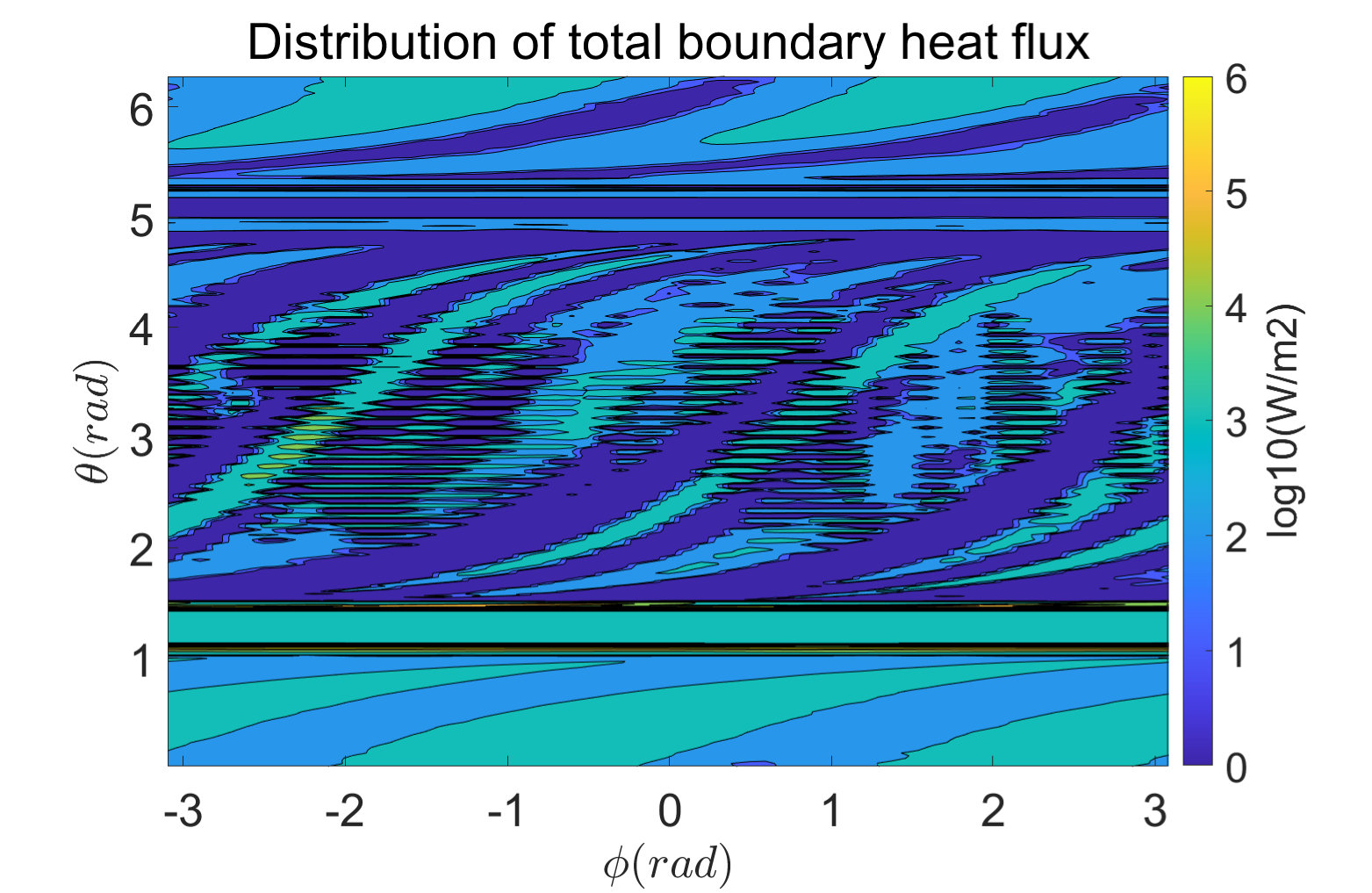}
}
\\
(c)&(d)
\etbl
\caption{The normal component of the combined conductive and convective heat flux of DH-FP-dt0 at times (a) $t=0.74ms$ and (b) $t=1.17ms$ across a surface $10cm$ within the simulation boundary, as well as that of DH-QP-dt0 at time (c) $t=1.09ms$ and (d) $t=1.90ms$. All four time slices coincide with peaks of MHD amplitude in their respective cases. The EQ-08 port locates approximately at $\gj=-2.57$ and the EQ-17 port locates at $\gj=0.57$.}
\label{fig:FWConHeatFlux_degraded_H}
\end{figure*}

The $\log_{10}$ of the normal component of the boundary heat flux at the two times when the MHD amplitude reaches its peak values for both cases are shown in Fig.\,\ref{fig:FWConHeatFlux_degraded_H}. Fig.\,\ref{fig:FWConHeatFlux_degraded_H}(a) and (b) correspond to two MHD peaks of DH-FP-dt0 and Fig.\,\ref{fig:FWConHeatFlux_degraded_H}(c) and (d) correspond to that of DH-QP-dt0. For DH-FP-dt0, a small asymmetry in the MHD response initially results in a weak drift motion and mild out going heat flux from the EQ-08 plume position at $\gj=-2.57$ and $t=0.74ms$ as is shown in Fig.\,\ref{fig:FWConHeatFlux_degraded_H}(a). Later on, at the time of the core relaxation, strong heat flux coming out from the X point of the $1/1$ magnetic island results in a local high pressure plasmoid close to the EQ-17 plume location at $\gj=0.57$ and $t=1.17ms$, which is accompanied by a strong outgoing heat flux shown in Fig.\,\ref{fig:FWConHeatFlux_degraded_H}(b). The heat flux associated with the second drift is comparable in amplitude with the heat fluxes in BH-FP-dt0 as is shown in Fig.\,\ref{fig:FWConHeatFlux}. With higher neon mixture ratio, DH-QP-dt0 found negligible plasmoid drift, and almost no accompanying outgoing heat flux even during the MHD peak, as is shown in  Fig.\,\ref{fig:FWConHeatFlux_degraded_H}(c) \& (d). This is responsible for the difference in the radiation fraction between BH-FP-dt0 and DH-QP-dt0 shown in Fig.\,\ref{fig:RadFraction}.

To summarize, the arrival of the vanguard fragments on the baseline H-mode pedestal would trigger strong outward plasmoid drift which is accompanied by edge stochastization, resulting in significant density expulsion and a large heat flux onto the wall. This is detrimental to the TQ mitigation. Considering the confinement degradation in the precursor phase as well as using higher neon concentration pellets can effectively mitigate such drift motion and the associated heat flux, thus increasing the radiation fraction.

\section{Radiation heat flux and its energy impact after neon dual-SPIs}
\label{s:HeatFlux}

Another important issue that has to be addressed is the asymmetry of the radiation heat deposition onto the ITER first wall. Concerns have been raised that if there exists strong asymmetry in the volumetric radiation power density, local parts of the first wall could still receive significant amount of heat flux leading to surface melting of PFCs. For the design of the ITER in-vessel components such as diagnostics or heating systems, the radiation peaking and the resulting heat fluxes to these components have been specified. However, validation of these specifications is still outstanding and the simulations presented here contribute to this validation process. To investigate such radiation heat deposition, we utilize the IMAS integrated RaySect/CHERAB code suite to obtain the wall heat flux via ray-tracing the radiation power density obtained by JOREK onto the realistic ITER first wall tiles and the stainless steel armour plate of the diagnostic port windows. Once the heat flux distribution is obtained, the accumulated energy impact is calculated by the method detailed in Section \ref{ss:EnergyImpact}, which then is compared to the maximum tolerable energy impact limit to assess the potential damage to the PFCs. This serves as a first demonstration of the JOREK capability to provide input for future more detailed PFC heat damage investigations. It should be noted that directly using the high time resolution JOREK data to calculate the energy impact is impractical due to the enormous data size, so we used lower time resolution data with linear interpolation between time slices to carry out the time integration needed in Eq.\,(\rfq{eq:EnergyImpact}).

\begin{table*}
\centering
\noindent
\btbl{|c|c|c|c|c|c|}
\hline
Case \# & Assim. Ne &  Assim. H & $f_{rad}$ & $t_{TQ}$ ($90\%$-$20\%$) & Max. $\gD Q$\\
\hline
BH-FP-dt0 & $9.91\times 10^{21}$  & $6.40\times 10^{23}$ & $49.4\%$ & $2.0ms$ & $25.4MWs^{1/2}/m^2$\\
\hline
DH-QP-dt0 & $2.53\times 10^{22}$  & $4.40\times 10^{23}$ & $89.8\%$ & $4.4ms$ & $5.7MWs^{1/2}/m^2$\\
\hline
DH-QP-dt1 & $2.15\times 10^{22}$  & $3.58\times 10^{23}$ & $86.3\%$ & $4.2ms$ & $14.9MWs^{1/2}/m^2$\\
\hline
\etbl
\caption{The neon and hydrogen assimilation, radiated fraction, TQ time and maximum local energy impact for three cases considered in this section.}
\label{tab:2}
\end{table*}

We mainly focus on three cases in this section: BH-FP-dt0, DH-QP-dt0 \& DH-QP-dt1 as shown in Table \ref{tab:1}. The overall assimilated neon, assimilated hydrogen within the LCFS, radiated energy fraction $f_{rad}$, approximate TQ timescale defined as the timescale of the thermal energy falling from $90\%$ to $20\%$ of the initial thermal energy, as well as the maximum local energy impact for those three cases are summarized in Table \ref{tab:2}. We are interested in DH-QP-dt0 \& DH-QP-dt1 since the quarter pellet dual-SPIs into the degraded H-mode shows good radiation fraction, as is demonstrated in Fig.\,\ref{fig:RadFraction}(b). We are interested in the comparison between the perfectly synchronized DH-QP-dt0 and the $1ms$ delayed DH-QP-dt1, since the mismatch between the arrival time of the fragment plumes means stronger asymmetry in the radiation power density and probably a larger maximum heat flux, and it is of interest to see if the $1ms$ delayed case would see local radiation energy impact exceeding the tolerable limit of the PFCs. The baseline H-mode BH-FP-dt0 is included to demonstrate the expected strongest radiation heat deposition onto the first wall due to the large radiation power, despite its smaller radiated fraction.

\begin{figure*}
\centering
\noindent
\btbl{c}
\parbox{3.5in}{
    \includegraphics[scale=0.3]{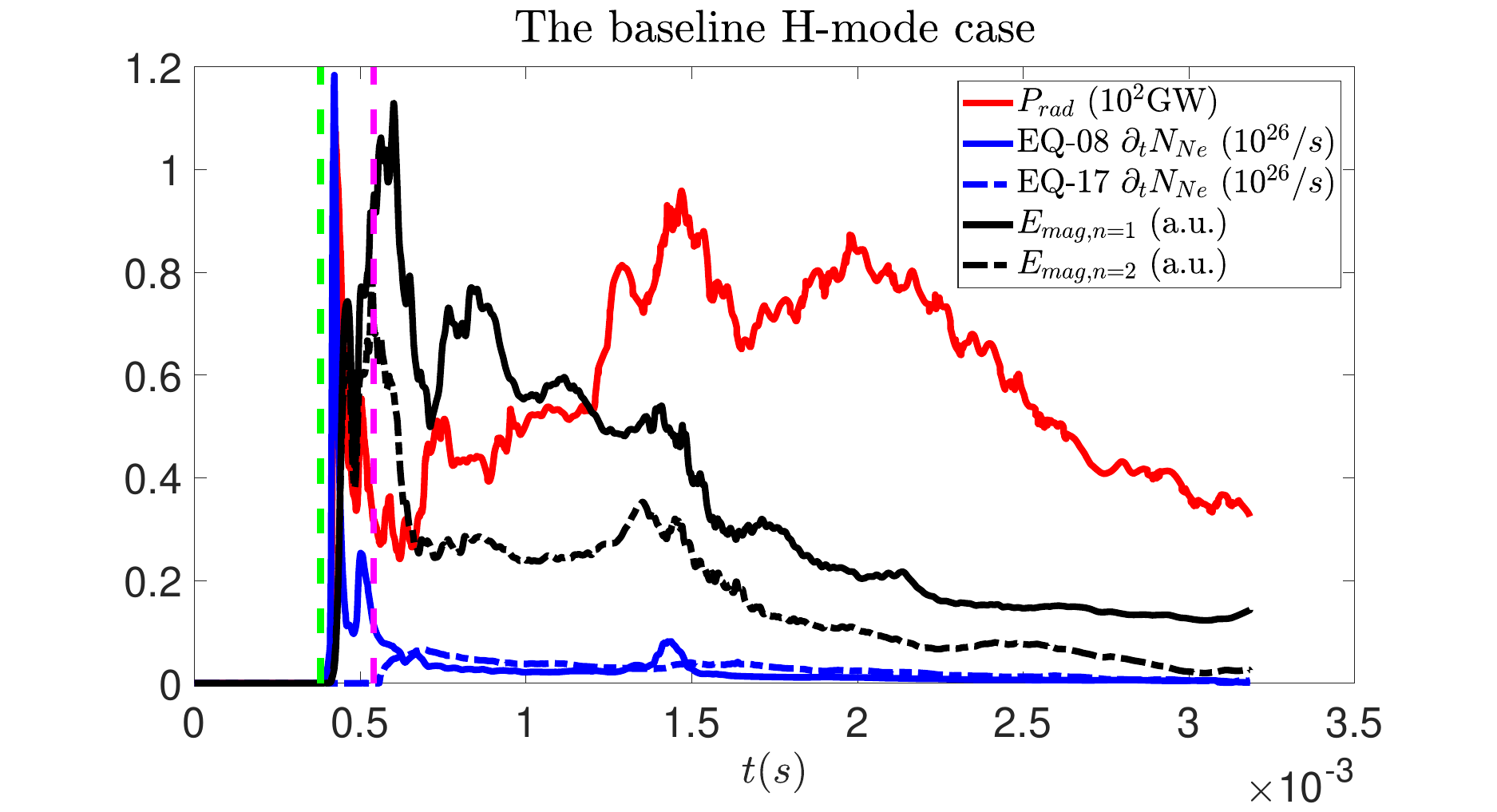}
}
\\
(a)
\\
\parbox{3.5in}{
    \includegraphics[scale=0.3]{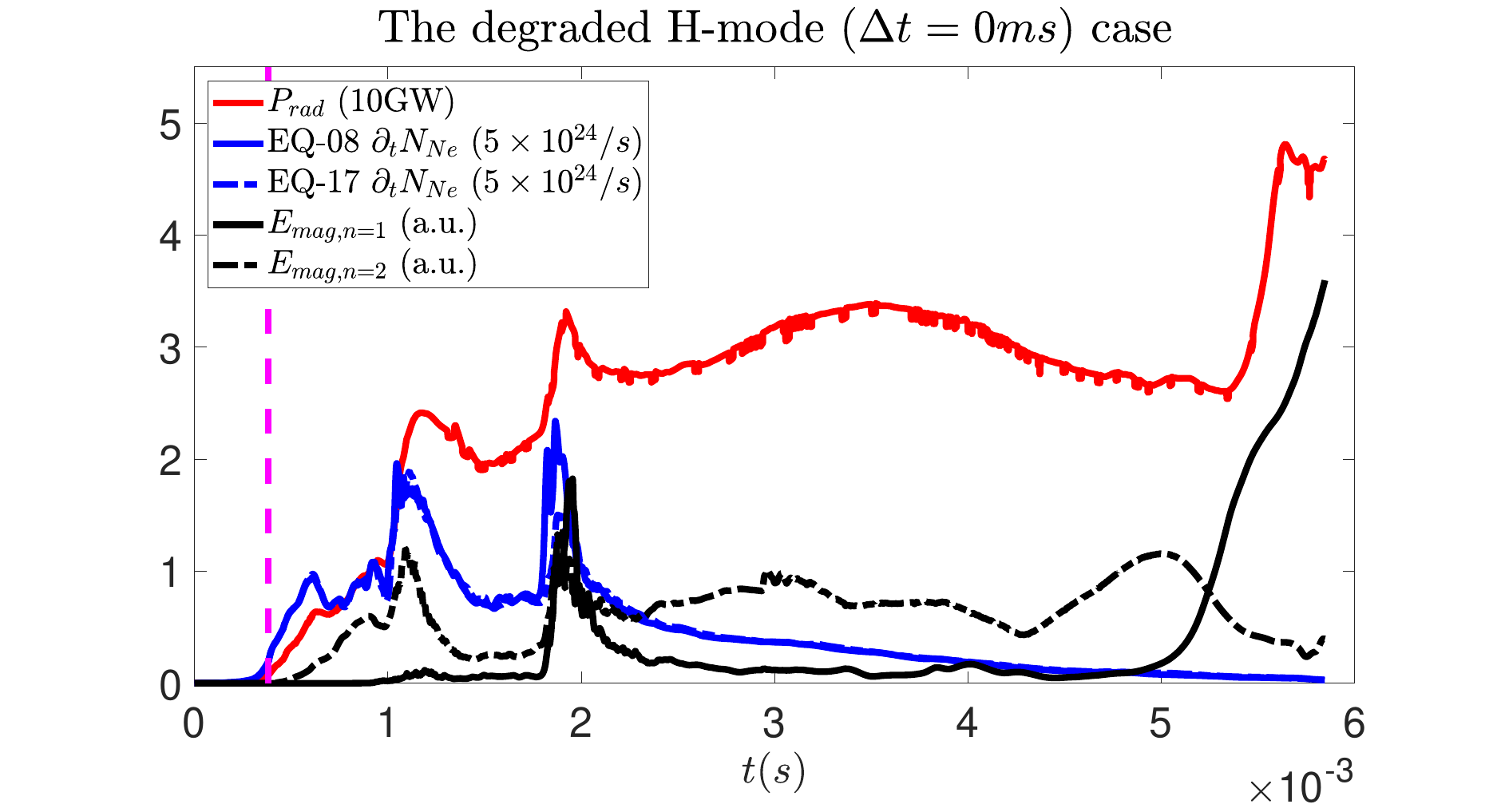}
}
\\
(b)
\\
\parbox{3.5in}{
    \includegraphics[scale=0.3]{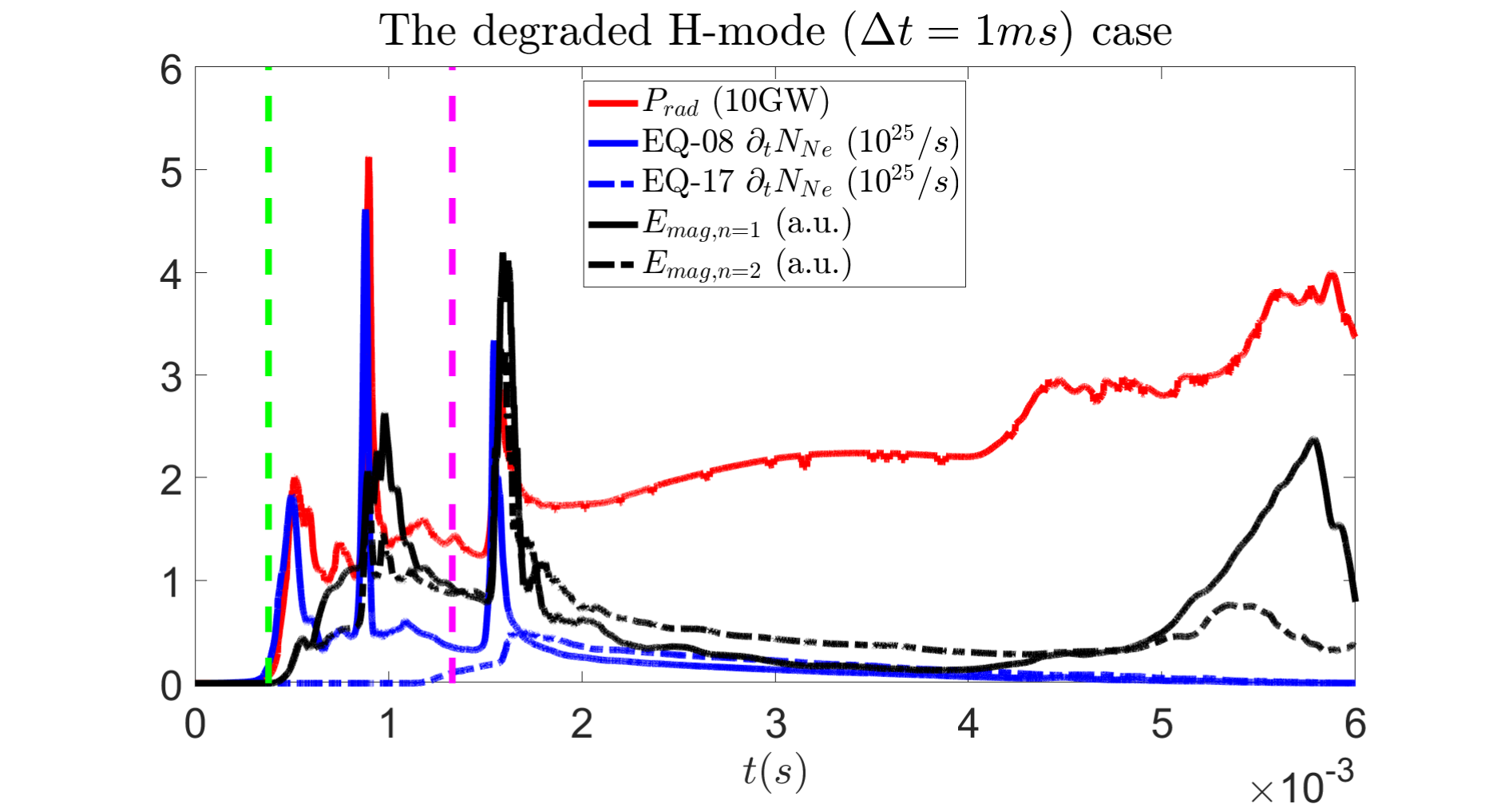}
}
\\
(c)
\etbl
\caption{The correlation between the radiation power, the neon ablation rate of both plumes, as well as the $n=1$ and $n=2$ magnetic energy perturbation for (a) BH-FP-dt0, (b)DH-QP-dt0 and (c)DH-QP-dt1. The approximate arrival time of the vanguard fragments on the equilibrium LCFS is represented by the green vertical dashed line for the EQ-08 plume and the magenta vertical dashed line for the EQ-17 plume. For the DH-QP-dt0 case, the two arrival time overlap with each other.}
\label{fig:Rad_Abl_E_mag}
\end{figure*}

\begin{figure*}
\centering
\noindent
\btbl{c}
\parbox{3.5in}{
    \includegraphics[scale=0.4]{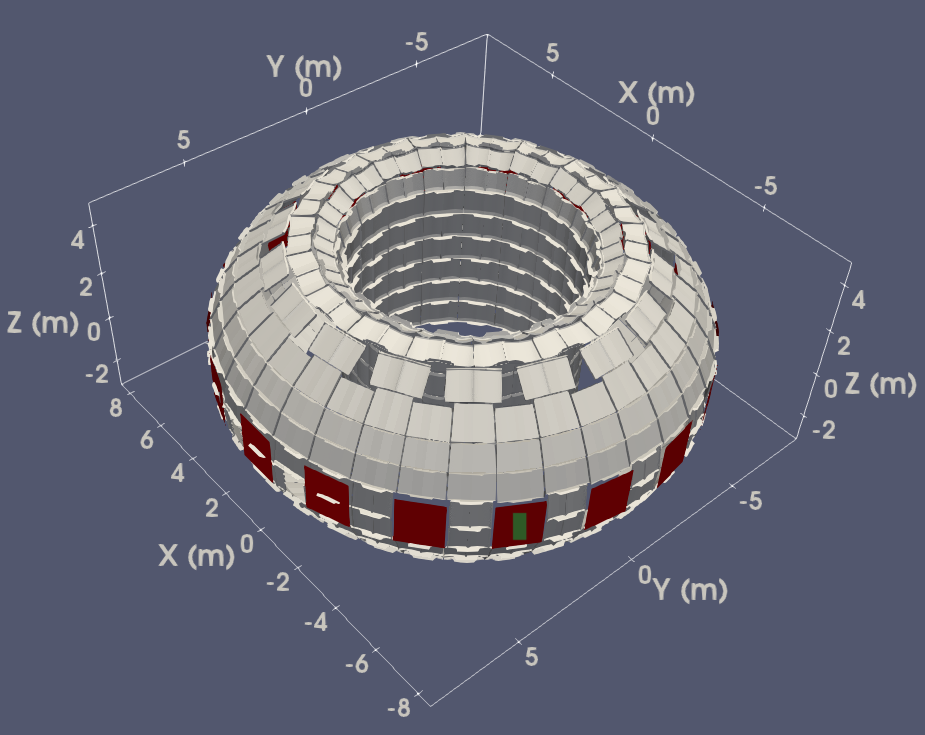}
}
\etbl
\caption{The ITER first wall (white) and the diagnostic port armour (red). The EQ-08 port locates at $X=-7.05m$ and $Y=4.57m$, it is the third port from right visible in the figure, marked by the green rectangle. The EQ-17 port locates at $X=7.45m$ and $Y=-3.84m$, on the other side of the torus.}
\label{fig:Wall}
\end{figure*}

\begin{figure*}
\centering
\noindent
\btbl{cc}
\parbox{2.4in}{
	\includegraphics[scale=0.2]{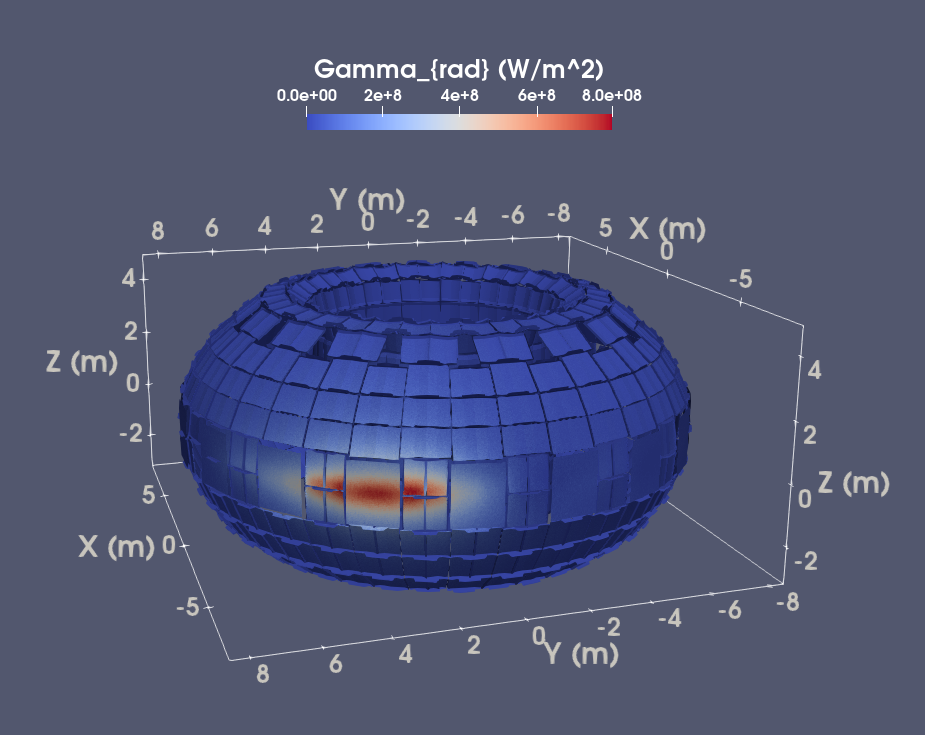}
}
&
\parbox{2.4in}{
	\includegraphics[scale=0.2]{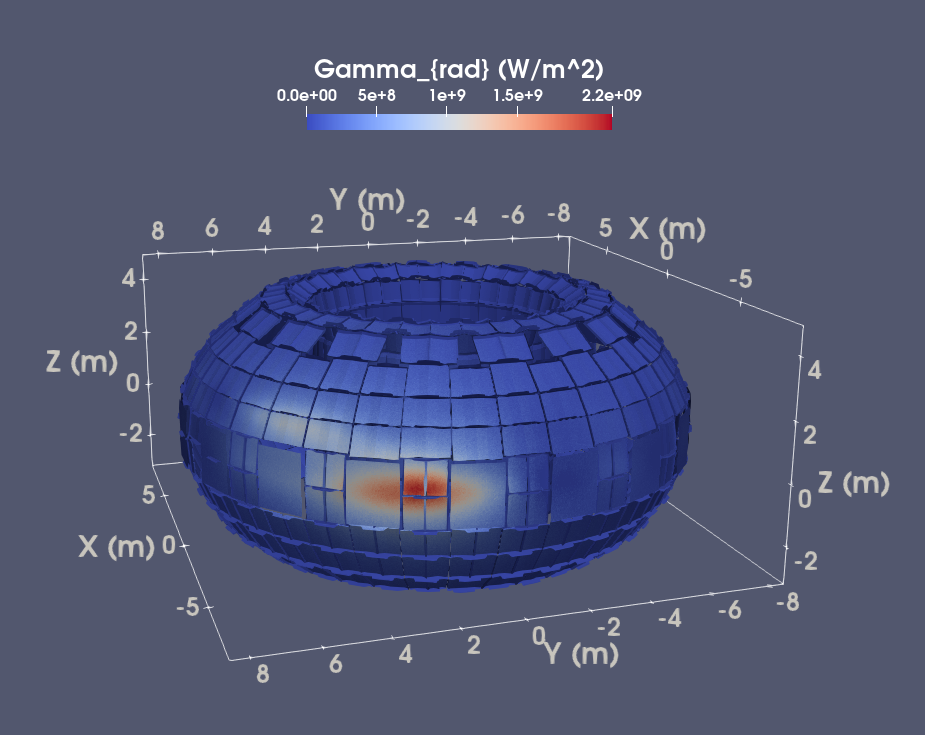}
}
\\
(a)&(b)
\\
\parbox{2.4in}{
	\includegraphics[scale=0.2]{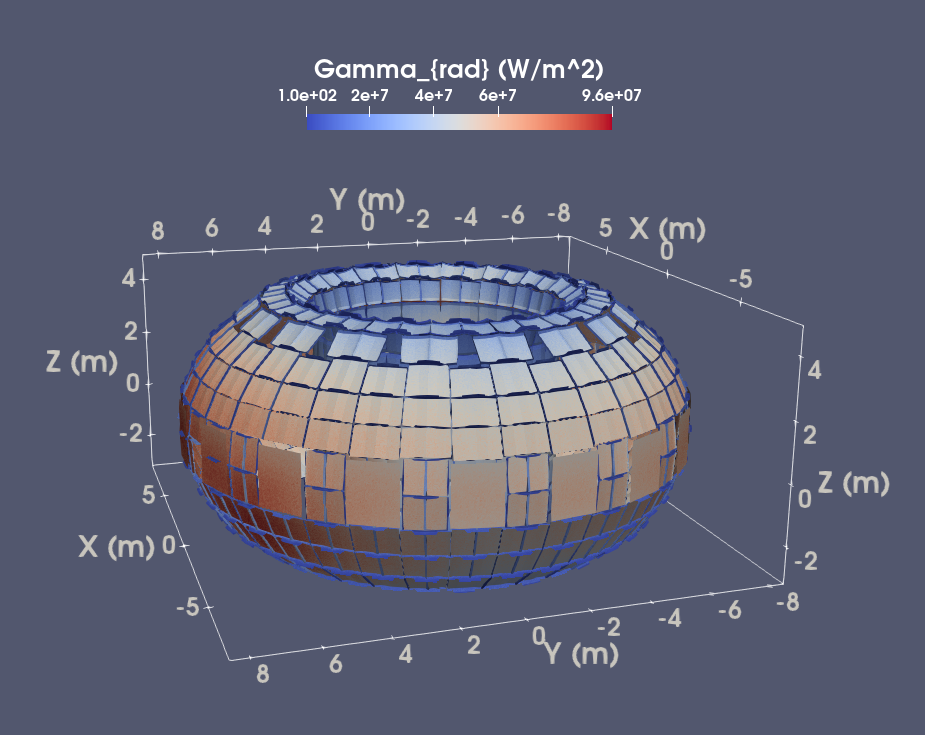}
}
&
\parbox{2.4in}{
	\includegraphics[scale=0.2]{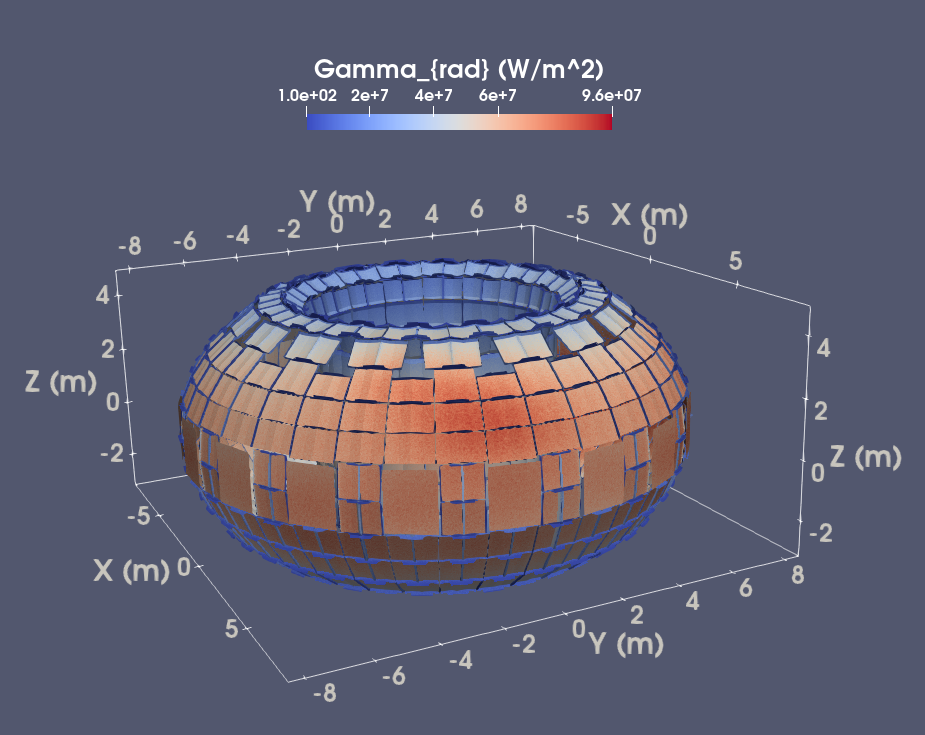}
}
\\
(c)&(d)
\etbl
\caption{The first wall heat flux of DH-QP-dt1 at (a) $t=0.52ms$ looking at port EQ-08, (b) $t=0.90ms$ looking at port EQ-08, (c) $t=5.74ms$ looking at port EQ-08 and (d) $t=5.74ms$ looking at port EQ-17. By the time of $t=5.74ms$, the radiation heat flux already becomes relaxed, and despite the high total radiation power, the radiation heat flux is low.}
\label{fig:HeatFlux_Degraded_H_1ms}
\end{figure*}

The correlation between the radiation power, the neon ablation rate of both plumes, as well as the $n=1$ and $n=2$ magnetic energy perturbation for all three cases are shown in Fig.\,\ref{fig:Rad_Abl_E_mag}, with red lines representing the radiation power, the blue solid lines representing the neon ablation rate of the EQ-08 plume, the blue chained lines representing that of the EQ-17 plume, the black solid lines representing the $n=1$ perturbed magnetic energy and the black chained lines representing the $n=2$ perturbed magnetic energy respectively. The approximate arrival time of the vanguard fragments on the equilibrium LCFS is represented by the green vertical dashed line for the EQ-08 plume and the magenta vertical dashed line for the EQ-17 plume. For BH-FP-dt0 shown in Fig.\,\ref{fig:Rad_Abl_E_mag}(a), the very sharp peak in ablation and radiation at approximately $t=0.42ms$ coincide with the strong plasmoid drift discussed in Section \ref{s:PlasmoidDrift}, the following peak in the MHD amplitude corresponds to the edge stochasticity shown in Fig.\,\ref{fig:DriftPoincare}. The later radiation peaks coincide with the core temperature relaxation and cooling, which corresponds a weaker MHD peak just before $t=1.5ms$. For DH-QP-dt0 shown in Fig.\,\ref{fig:Rad_Abl_E_mag}(b), there are two radiation peaks in the early phase of injection at approximately $t=1.1ms$ and $t=2ms$, the later peak corresponds to the core temperature relaxation. These peaks are closely correlated with peaks in ablation and MHD amplitude. This correlation is the result of very nonlinear interaction between the fragments and the MHD mode. The ablation creates locally peaked lowly charged impurity density profile, thus result in strong local radiation cooling. This local cooling and local pressure peak incurs helical current perturbation which destabilize resonant MHD modes. The growing MHD modes result in enhanced outgoing heat flux which further strengthen the ablation in the plume. Thus the strong correlation between the radiation, ablation and MHD amplitude peaks especially in the early injection phase. DH-QP-dt1 shows similar correlation as is shown in Fig.\,\ref{fig:Rad_Abl_E_mag}(c).

\begin{figure*}
\centering
\noindent
\btbl{cc}
\parbox{2.4in}{
	\includegraphics[scale=0.2]{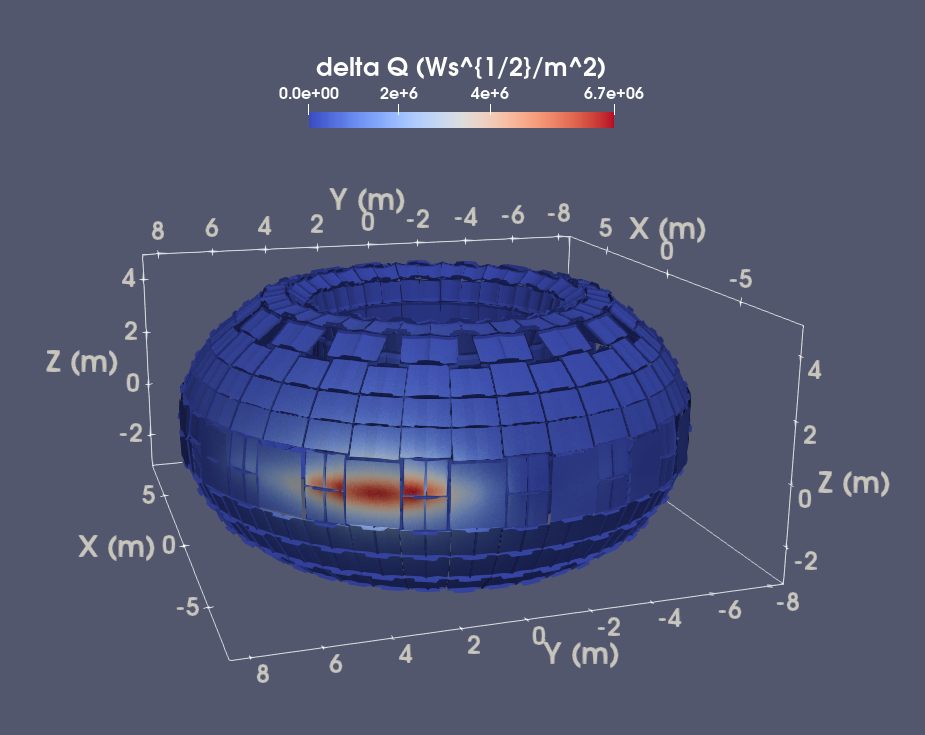}
}
&
\parbox{2.4in}{
	\includegraphics[scale=0.2]{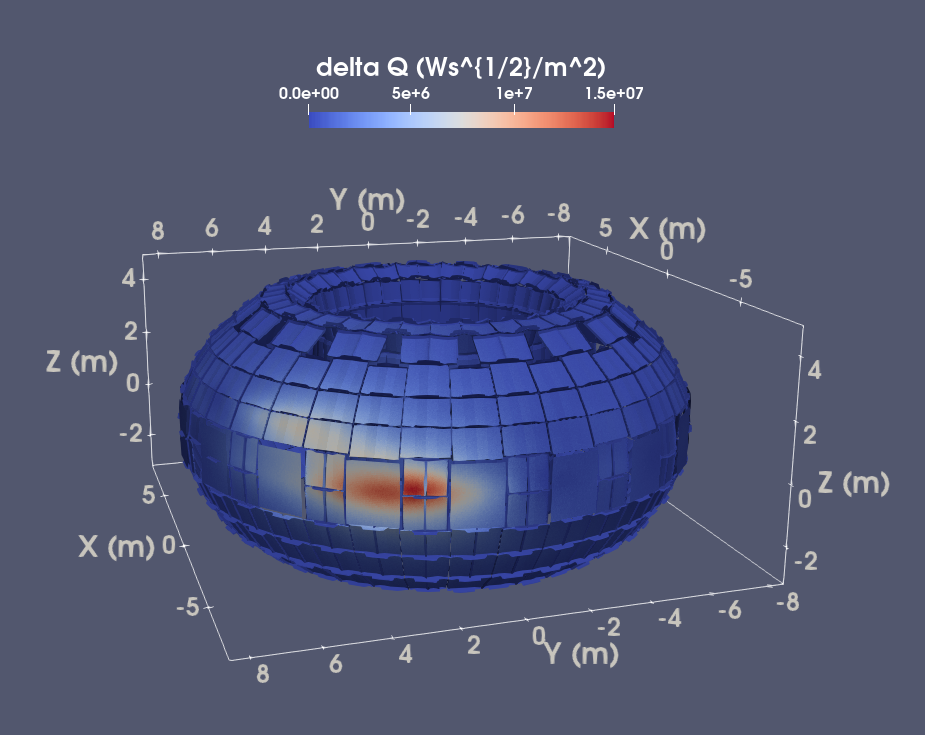}
}
\\
(a)&(b)
\\
\parbox{2.4in}{
	\includegraphics[scale=0.2]{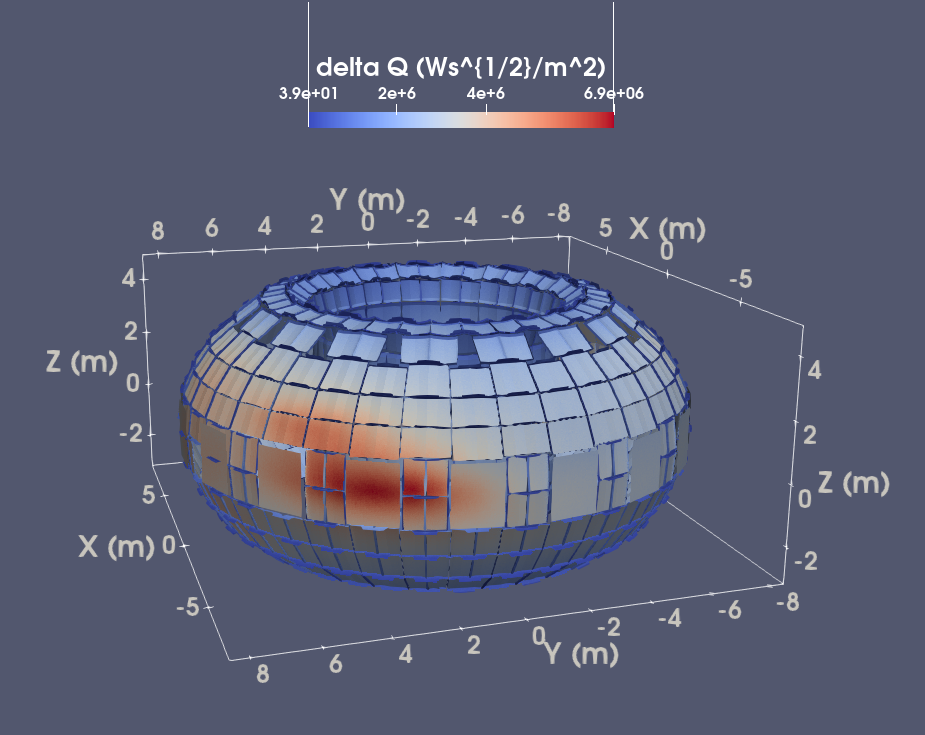}
}
&
\parbox{2.4in}{
	\includegraphics[scale=0.2]{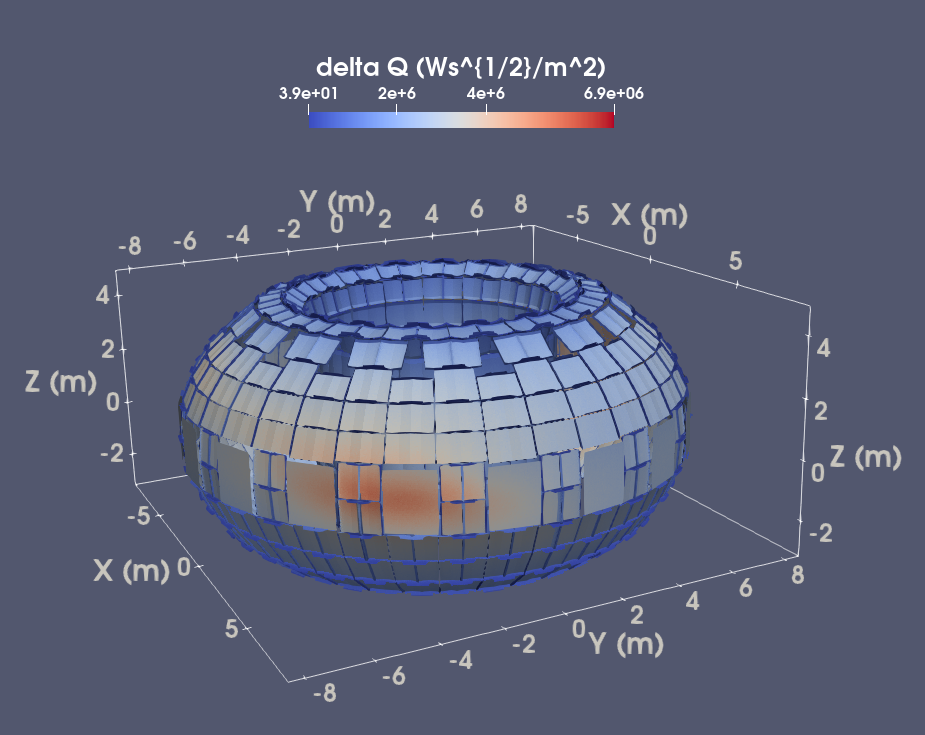}
}
\\
(c)&(d)
\etbl
\caption{The first wall energy impact of DH-QP-dt1 calculated using Eq.\,\rfq{eq:EnergyImpact} at (a) $t=0.56ms$ looking at port EQ-08, (b) $t=0.94ms$ looking at port EQ-08, (c) $t=5.74ms$ looking at port EQ-08 and (d) $t=5.74ms$ looking at port EQ-17. }
\label{fig:EnergyImpact_Degraded_H_1ms}
\end{figure*}

For each case, a few time slices are chosen to show both the instantaneous heat flux and the accumulated energy impact onto the first wall. The time around the radiation peaks shown in Fig.\,\ref{fig:Rad_Abl_E_mag} are especially of interest, so we choose the time slices to be at the radiation peak for the heat flux analysis and right after the peak for the energy impact analysis for the early injection phase. For completion, a few time slices are also chosen for the late injection phase when the radiation power density distribution is already relaxed within the plasma. The ITER first wall tiles as well as the diagnostic window armour plates are shown in Fig.\,\ref{fig:Wall}, where the tiles are shown in white and the armour plates are shown in red. The EQ-08 port locates at $X=-7.05m$ and $Y=4.57m$, it is the third port from right visible in the figure. The EQ-17 port locates at $X=7.45m$ and $Y=-3.84m$, on the other side of the torus.

\begin{figure*}
\centering
\noindent
\btbl{cc}
\parbox{2.4in}{
	\includegraphics[scale=0.2]{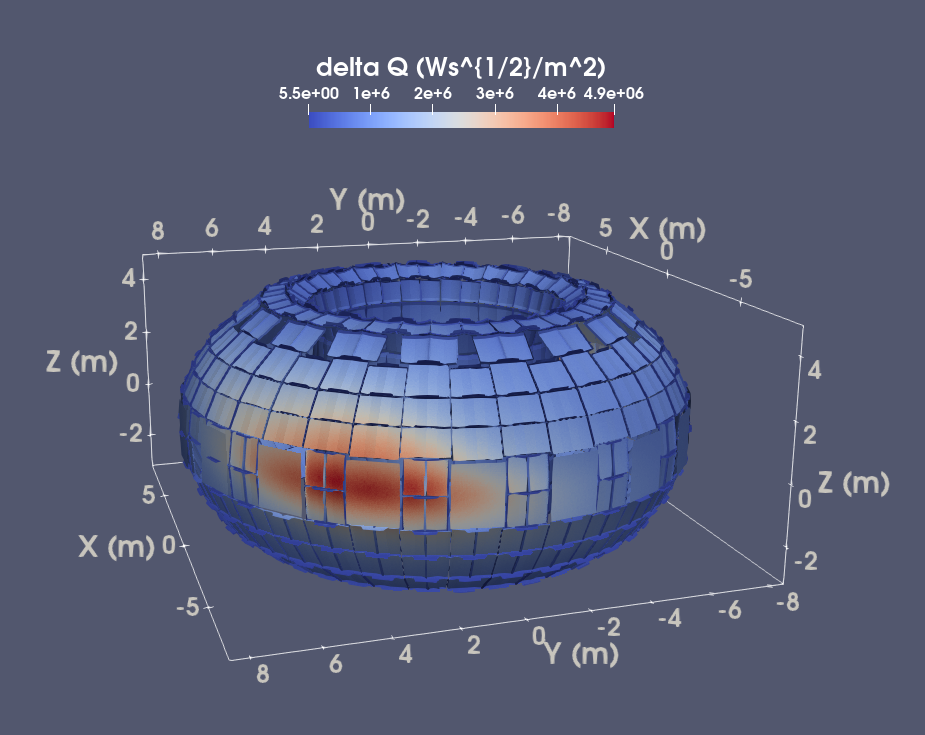}
}
&
\parbox{2.4in}{
	\includegraphics[scale=0.2]{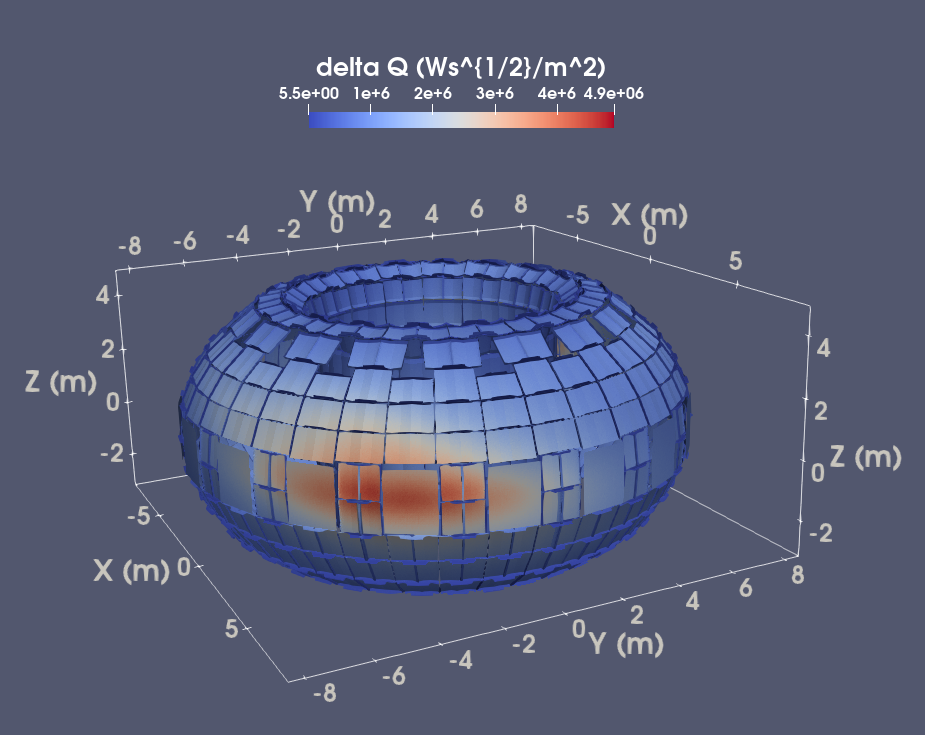}
}
\\
(a)&(b)
\\
\parbox{2.4in}{
	\includegraphics[scale=0.2]{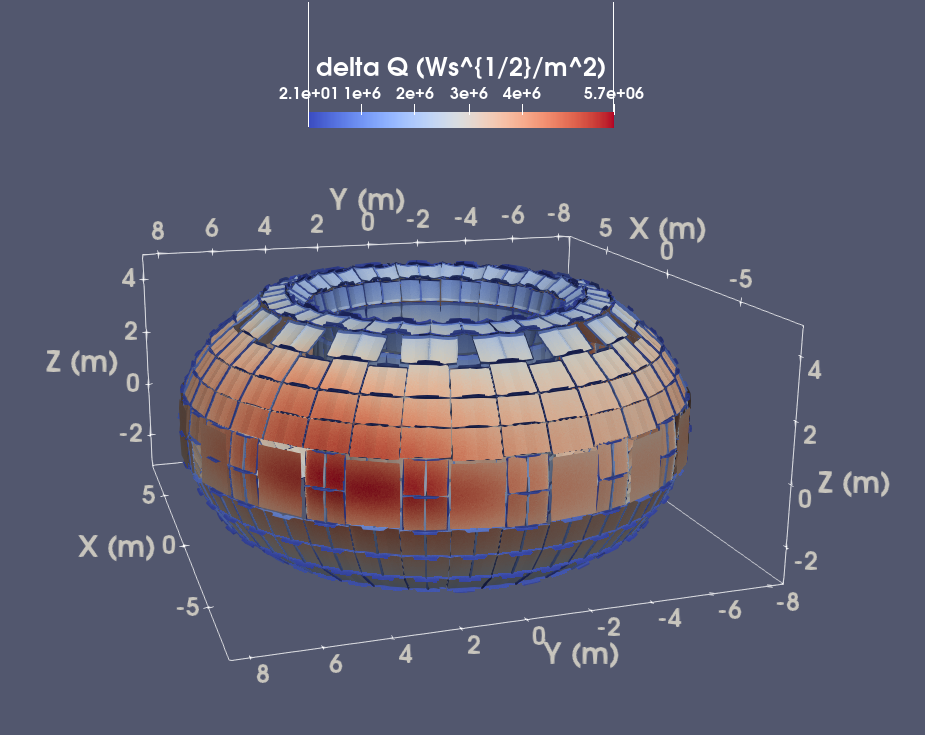}
}
&
\parbox{2.4in}{
	\includegraphics[scale=0.2]{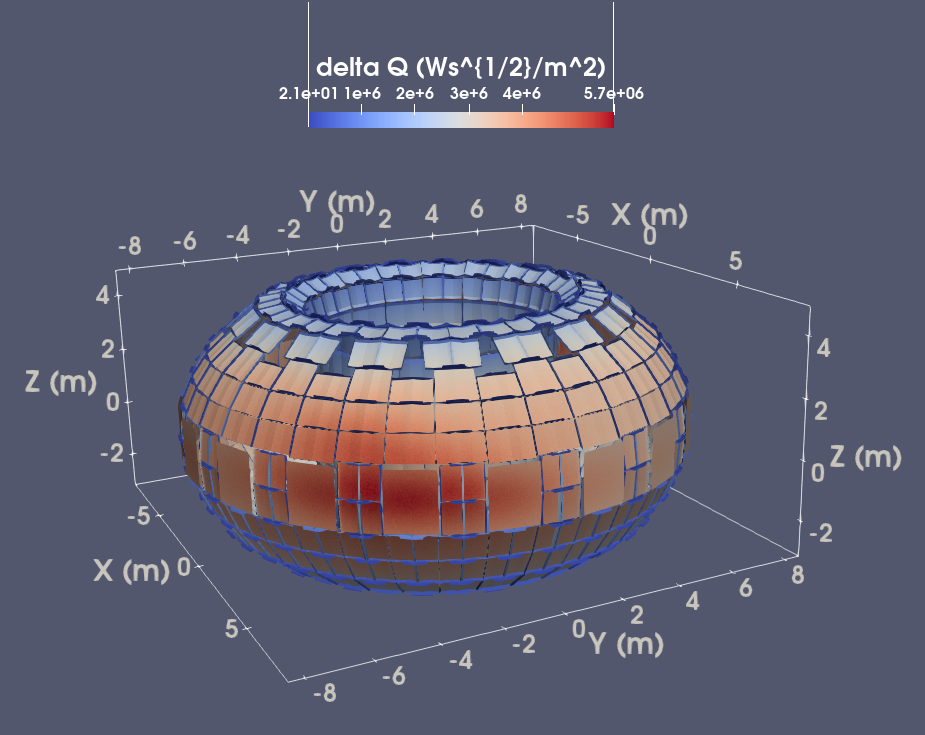}
}
\\
(c)&(d)
\etbl
\caption{The first wall energy impact of DH-QP-dt0 calculated using Eq.\,\rfq{eq:EnergyImpact} at (a) $t=2.06ms$ looking at port EQ-08, (b) $t=2.06ms$ looking at port EQ-17, (c) $t=5.64ms$ looking at port EQ-08 and (d) $t=5.64ms$ looking at port EQ-17.}
\label{fig:EnergyImpact_Degraded_H_0ms}
\end{figure*}

We first take a look at DH-QP-dt1. The instantaneous heat flux onto the first wall is shown in Fig.\,\ref{fig:HeatFlux_Degraded_H_1ms}. Since the EQ-08 plume arrives earlier in this case, the radiation heat flux concentrates around the EQ-08 port in the early injection phase as can be seen from Fig.\,\ref{fig:HeatFlux_Degraded_H_1ms}(a) and (b), at $t=0.52ms$ and $t=0.90ms$ respectively. These two times correspond to the first two peaks in the radiation power shown in Fig.\,\ref{fig:Rad_Abl_E_mag}(c). The toroidal elongation of the heat spots in Fig.\,\ref{fig:HeatFlux_Degraded_H_1ms}(a) and (b) is due to the toroidally elongated density source shape as we have discussed in Section \ref{ss:System}. It can be seen that initially the radiation heat flux is almost symmetric around the EQ-08 port in Fig.\,\ref{fig:HeatFlux_Degraded_H_1ms}(a), but shifts to the right in Fig.\,\ref{fig:HeatFlux_Degraded_H_1ms}(b). This is a result of the interplay between the impurity density distribution and the outgoing heat flux which moves the radiation power density peak away from the fragment location that has been reported previously \cite{Di2021NF}. Ultimately, the radiation power density relaxes over the plasma volume both due to the impurity density relaxation and the arrival of the second plume. As a result, the first wall radiation heat flux becomes more or less uniform in the late phase of the SPI, as can be seen in Fig.\,\ref{fig:HeatFlux_Degraded_H_1ms}(c) and (d) which show the heat flux around both port EQ-08 and EQ-17 at $t=5.74ms$. At this time, despite the high total radiation power as is shown in Fig.\,\ref{fig:Rad_Abl_E_mag}(c), the maximum radiation heat flux onto the first wall is one order of magnitude lower compared with the early injection phase shown in Fig.\,\ref{fig:HeatFlux_Degraded_H_1ms}(a) and (b). Another notable feature is that the heat flux on the armour plate is smaller compared with that of the first wall tiles, which is due to a geometric effect.

\begin{figure*}
\centering
\noindent
\btbl{cc}
\parbox{2.4in}{
	\includegraphics[scale=0.2]{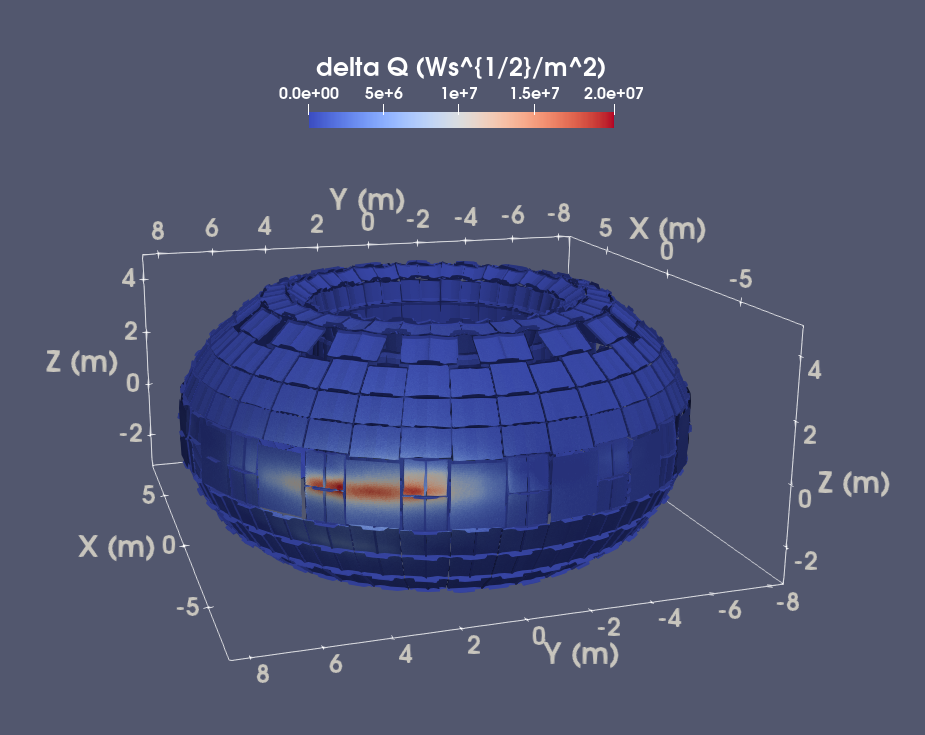}
}
&
\parbox{2.4in}{
	\includegraphics[scale=0.2]{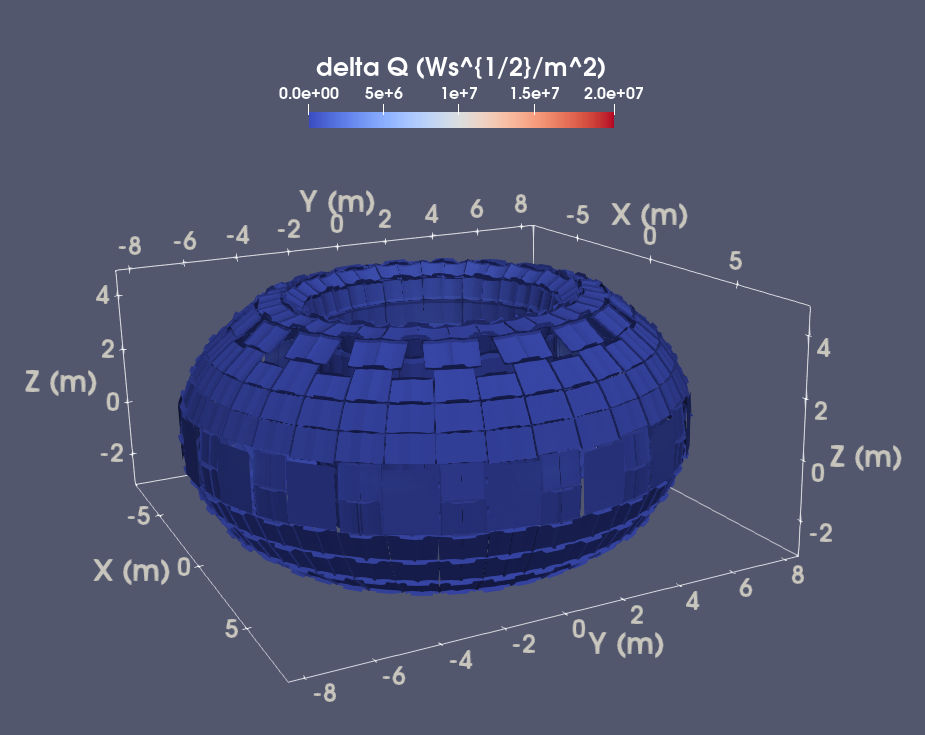}
}
\\
(a)&(b)
\\
\parbox{2.4in}{
	\includegraphics[scale=0.2]{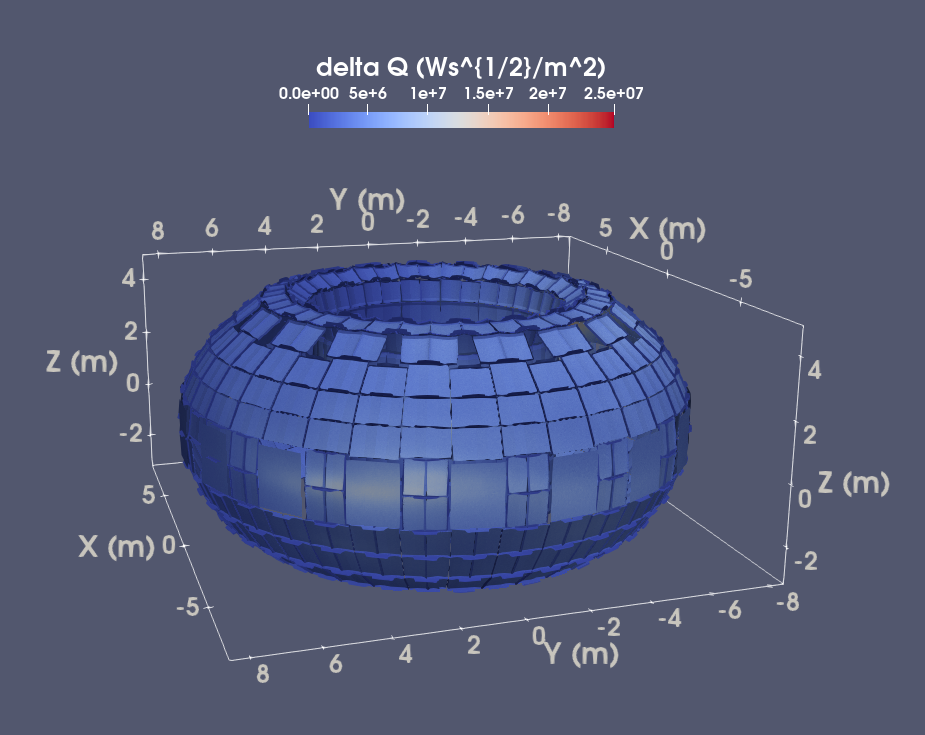}
}
&
\parbox{2.4in}{
	\includegraphics[scale=0.2]{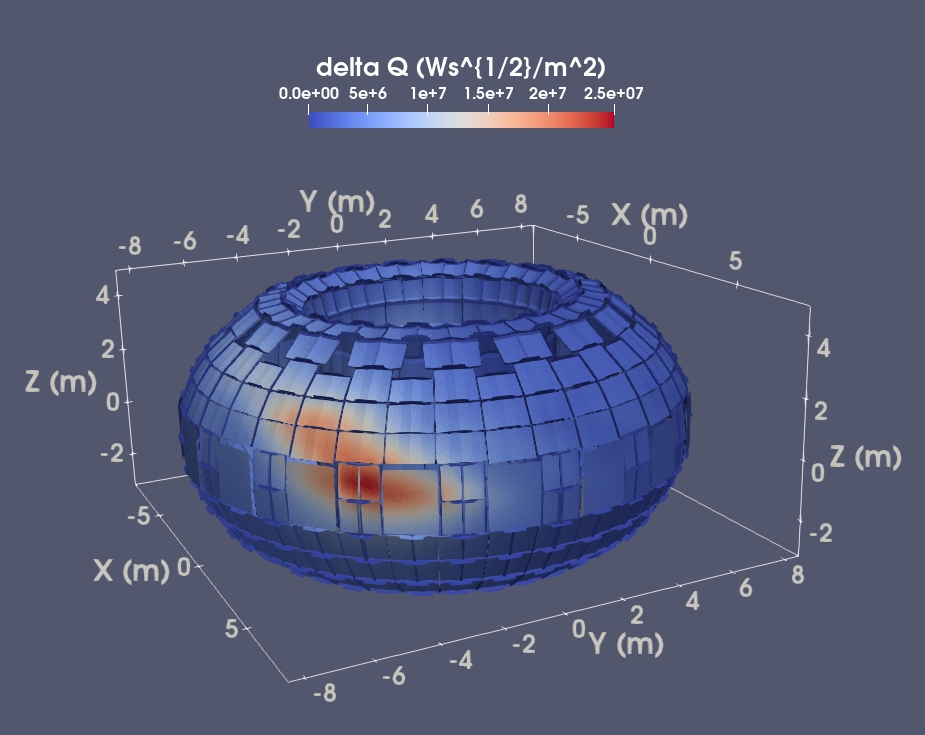}
}
\\
(c)&(d)
\etbl
\caption{The first wall energy impact of BH-FP-dt0 calculated using Eq.\,\rfq{eq:EnergyImpact} at (a) $t=0.43ms$ looking at port EQ-08, (b) $t=0.43ms$ looking at port EQ-17, (c) $t=2.00ms$ looking at port EQ-08 and (d) $t=2.00ms$ looking at port EQ-17.}
\label{fig:EnergyImpact_Full_H}
\end{figure*}

The energy impact factor at several chosen times for the same case is shown in Fig.\,\ref{fig:EnergyImpact_Degraded_H_1ms}. Fig.\,\ref{fig:EnergyImpact_Degraded_H_1ms}(a) and (b) show the energy impact just after the radiation spike at $t=0.56ms$ and $t=0.94ms$, corresponding to Fig.\,\ref{fig:HeatFlux_Degraded_H_1ms}(a) and (b), and indeed one can see the energy impact distribution correlates well with that of the transient heat flux. At the end of the second radiation spike at $t=0.94ms$, part of the stainless steel armour plate shows the energy impact approaching the maximum tolerable limit of $13MJs^{-1/2}/m^2$ \cite{Pitts2015JNM}. Another feature is that the tiles beside the EQ-08 plate show higher energy impact, consistent with the heat flux behaviour shown in \ref{fig:HeatFlux_Degraded_H_1ms}(b). Nevertheless, the energy impact on those tiles is well below the maximum tolerable limit for the tungsten at $38MJs^{-1/2}/m^2$ \cite{Herrmann2002PPCF}. In the late injection phase at $t=5.74ms$, despite the high total radiation power, the energy impact onto the first wall is much more uniform and stays well below the stainless steel limit, as is shown in \ref{fig:EnergyImpact_Degraded_H_1ms}(c) and (d) which look at the EQ-08 and EQ-17 port respectively.

As a comparison, the first wall energy impact in the early and late injection phase for DH-QP-dt0 is shown in  Fig.\,\ref{fig:EnergyImpact_Degraded_H_0ms}. The energy impact at $t=2.06ms$, after the second radiation peak shown in Fig.\,\ref{fig:Rad_Abl_E_mag}(b), around the EQ-08 and EQ-17 ports is shown in Fig.\,\ref{fig:EnergyImpact_Degraded_H_0ms}(a) and (b) respectively, while the late time energy impact at $t=5.64ms$ is shown in Fig.\,\ref{fig:EnergyImpact_Degraded_H_0ms}(c) and (d) for both locations. Due to the good synchronization between the injectors, both ports see similar energy impact accumulation in the early injection phase, and the maximum energy impact stays well below that of the stainless steel limit, let alone the tungsten one. In the late injection phase, due to the relaxation of the radiation power density, the energy impact becomes rather uniform across the torus apart from some poloidal asymmetry due to the shape of the first wall cross-section, as can be seen in Fig.\,\ref{fig:EnergyImpact_Degraded_H_0ms}(c) and (d). Again, despite the higher radiation power, the maximum energy impact stays well below the tolerable limit of the stainless steel.

Finally, we take a look at BH-FP-dt0 to demonstrate the highest radiation energy impact in our simulations. This high energy impact is because of both the high total radiation power and the asymmetry in the fragment plume we explained previously. Due to this asymmetry, we expect the radiation to concentrate around the EQ-08 port since the vanguard of its plume would first reach the plasma and trigger the plasmoid drift motion discussed in Section \ref{s:PlasmoidDrift}. Indeed, as is shown in Fig.\,\ref{fig:EnergyImpact_Full_H}(a) and (b) at $t=0.43ms$, after the sharp radiation peak caused by the vanguard fragments arriving on the baseline H-mode pedestal, significant energy impact is accumulated around the EQ-08 port while the energy impact around the EQ-17 port is negligible. The energy impact of this radiation pulse already exceeds the tolerable limit of the stainless steel, thus we expect to see melting of the EQ-08 armour plate even without considering the conductive/convective heat flux accompanying the plasmoid drift. The maximum radiation energy impact is still well below the tungsten limit, however. As time goes on and the fragments fly further inward, we see the radiation asymmetry turns to the other side of the torus around the EQ-17 port, as is shown in Fig.\,\ref{fig:EnergyImpact_Full_H}(c) and (d) at $t=2.00ms$. The radiation energy impact also concentrates around the injection port at this time, but shows stronger helical feature, representing the parallel relaxation of the impurity density. Again, the maximum energy impact exceeds the stainless steel limit, but stays well below the tungsten one.

\section{Conclusion and discussion}
\label{s:Conclusion}

JOREK simulations with collisional-radiative impurities are carried out for dual-SPIs into ITER H-mode plasma to investigate the heat deposition during disruption mitigations. Two sets of equilibria are considered, one baseline H-mode following the default ITER H-mode scenario, and one degraded H-mode where an artificial H-L back transition is induced to mimic the confinement degradation in the disruption precursor phase. Into these two equilibria, two sets of dual-SPI configuration are simulated, the full pellet case with small neon fraction and the quarter pellet case where the neon amount is the same but the hydrogen amount is reduced to one quarter of the full pellet case.

It is found that with baseline H-mode and full pellet, a large local plasmoid over-pressure occurs near the vanguard fragments, which induces polarization and consequential drift motion outward along the major radius. This drift motion causes evident density expulsion and is accompanied by edge stochasticity and strong outward heat flux, both detrimental to the disruption mitigation. The transient heat flux is so large that even with expose time on the order of $1\gm s$, the resulting energy impact could still be on the order of $10^8MJs^{-1/2}/m^2$, exceeding the tolerable limit of the tungsten. By considering the pre-disruption confinement degradation and using the higher neon fraction quarter pellet, such local over-pressure could be reduced, thus mitigating the drift motion and the accompanying heat flux. In these scenarios, the transient heat flux is significantly diminished and the radiation fraction during the TQ is improved.

The radiation heat flux and its accumulated energy impact onto the first wall is obtained by ray-tracing from the volumetric radiation power density distribution provided by JOREK. Three cases are investigated: the baseline H-mode with full pellet case and two degraded H-mode with quarter pellet cases with perfectly synchronization and $1ms$ delay between the injectors respectively. It is found that the radiation heat deposition tends to concentrate around the injecting port in the early injection phase, while it spreads over larger areas towards the end of the TQ. With the quarter pellet, degraded H-mode and $1ms$ injection delay, the maximum radiation energy impact approaches the tolerable limit of the stainless steel plate while staying far below the tungsten limit. Good synchronization is found to further relax the energy impact distribution and reduce its maximum value, such that it never reaches the stainless steel limit for the perfect synchronized degraded H-mode case. These results are reassuring since they suggest the degraded H-mode with quarter pellet case could achieve high radiation fraction without the risk of melting the first wall. In the baseline H-mode case, despite the strong radiation power, its energy impact still stays well below the tungsten limit, although it exceeds that of the stainless steel substantially, hence one would expect melting of the diagnostic window armour plates. However, even if such melting occurs, it would not impede continued operation of ITER, as it has been found in previous studies with $22MJs^{-1/2}/m^2$ heat pulses that the transient surface melting would result in a surface roughening increasing at a rate of 1-2 micrometers per pulse, but no loss of material \cite{Klimov2013JNM}. Another interesting observation is that the maximum energy impact is approaching the melting limit of the beryllium at $28MJs^{-1/2}/m^2$ \cite{Pitts2015JNM}, so that the new ITER tungsten tiles provide more margin against radiation heat deposition compared with previous beryllium ones.

A few assumptions in our investigations might impact our results presented above. First of all is our toroidally elongated ablation density source, which result in enhanced impurity density spreading along the toroidal direction. One could expect that a more realistic, and expensive, $\gD\gj_{NG}$ would induce stronger impurity density peak and radiation power density peak, thus more localized radiation heat flux. This would especially be important if the radiating source is very close to the wall as in the early injection phase, or if the plasmoid drift causes the radiating plasmoid to move outward towards the wall. Another consequence of this elongated toroidal density source is that it tends to result in less concentrated pressure peak since the density is more spread out, thus reducing the drive for the plasmoid drift. It is however hard to say if a more realistic toroidal elongation would result in stronger drift, as the accompanying stronger radiation would tend to reduce the local pressure, thus counter-act the pressure peaking effect discussed above. This is left for future studies.
Furthermore, we also did not consider the opacity of the plasma. The high density plasmoid around the fragment plume could absorb significant portion of the radiation and re-emit them at a later time \cite{Aleynikov2024NF}, reducing the radiation peak. Resolving these two competing factors is part of our future plan.
Last, our choice of viscosities might impact the TQ duration, as is demonstrated in Ref\cite{McClenaghan2023NF}. The current values we chose are the minimum value we could use without rampant numerical instabilities. Using smaller viscosity might results in more rapid MHD growth and quicker TQ. However, as is shown in Ref\cite{McClenaghan2023NF}, changing the viscosity from $2000m^2/s$ to $500m^2/s$ only reduces the TQ duration from $2.9ms$ to $2.6ms$ in their viscosity scan. Hence we don't expect drastic changes to our conclusion should a more realistic viscosity is used.

Nevertheless, our study reported above serves as a first look into the radiation heat deposition on the ITER first wall during its H-mode TQ mitigation process using non-linear 3D simulation, which is part of ongoing validation work for the ITER PFC heat load specification. This study also provides input for more detailed heat transport and material damage studies in the future.

\vskip1em
\centerline{\bf Acknowledgments}
\vskip1em

  The authors thank Y. Yuan, L. Cheng and A. Matsuyama for fruitful discussion. ITER is the Nuclear Facility INB no. 174. The views and opinions expressed herein do not necessarily reflect those of the ITER Organization. This publication is provided for scientific purposes only. Its contents should not be considered as commitments from the ITER Organization as a nuclear operator in the frame of the licensing process. Part of this work is supported by the National MCF Energy R\&D Program of China under Grant No. 2019YFE03010001. This work has been co-funded by the ITER Organization under the implementing agreement IO/IA/19/4300002053. This work was supported in part by the Swiss National Science Foundation. Part of this work has been carried out within the framework of the  EUROfusion Consortium, funded by the European Union via the Euratom  Research and Training Programme (Grant Agreement No. 101052200 - EUROfusion). The Swiss contribution to this work has been funded by the Swiss State Secretariat for Education, Research and Innovation (SERI). Views and opinions expressed are however those of the  author(s) only and do not necessarily reflect those of the European  Union, the European Commission or SERI. Neither the European Union nor the  European Commission nor SERI can be held responsible for them. This work is carried out partly on Tianhe-3 operated by NSCC-TJ and partly on the ITER cluster.


\vskip1em
\centerline{\bf References}
\vskip1em

\begin{thebibliography}{99}

\bibitem{Lehnen2015JNM}
M. Lehnen, K. Aleynikova, P.B. Aleynikov et al., ``Disruptions in ITER and strategies for their control and mitigation'' J. Nucl. Mater., {\bf 463}, 39-48 (2015).

\bibitem{Li2020NF}
Y. Li, Z.Y. Chen, W. Yan et al., ``Comparison of disruption mitigation from shattered pellet injection with massive gas injection on J-TEXT'', Nucl. Fusion {\bf 61} 126025 (2020).

\bibitem{Sheikh2021NF}
U.A. Sheikh, D. Shiraki, R. Sweeney et al., ``Disruption thermal load mitigation with shattered pellet injection on the Joint European Torus (JET)'', Nucl. Fusion {\bf 61} 126043 (2021).

\bibitem{Jachmich2022NF}
S. Jachmich, U. Kruezi, M. Lehnen et al., ``Shattered pellet injection experiments at JET in support of the ITER disruption mitigation system design'', Nucl. Fusion {\bf 62} 026012 (2022).

\bibitem{Jang2022FED}
Juhyeok Jang, Jayhyun Kim, Jaewook Kim et al., ``Radiation distribution for shattered pellet injection experiment with AXUV array diagnostics in KSTAR'', Fusion Eng. Des. {\bf 180} 113172 (2022).

\bibitem{Stein-Lubrano2024NF}
B. Stein-Lubrano, R. Sweeney, D. Bonfiglio et al., ``3D Radiated Power Analysis of JET SPI Discharges Using the Emis3D Forward Modeling Tool'', Nucl. Fusion accepted (2024).

\bibitem{Kim2019POP}
C. C. Kim, Y. Liu, P. B. Parks et al., ``Shattered pellet injection simulations with NIMROD'' Phys. Plasmas, {\bf 26}, 042510 (2019).

\bibitem{Di2021NF}
D. Hu, E. Nardon, M. Hoelzl et al., ``Radiation asymmetry and MHD destabilization during the thermal quench after impurity Shattered Pellet Injection'' Nucl. Fusion, {\bf 61}, 026015 (2021).

\bibitem{Di2023NF}
D. Hu, E. Nardon, F.J. Artola et al., ``Collisional-radiative simulation of impurity assimilation, radiative collapse and MHD dynamics after ITER Shattered Pellet Injection'' Nucl. Fusion, {\bf 63}, 066008 (2023).

\bibitem{McClenaghan2023NF}
J. McClenaghan, B.C Lyons, C.C. Kim et al., ``MHD modeling of shattered pellet injection in JET'' Nucl. Fusion, {\bf 63}, 066029 (2023).

\bibitem{Herrmann2002PPCF}
A. Herrmann, ``Overview on stationary and transient divertor heat loads'' Plasma Phys. Control. Fusion {\bf 44} (2002) 883–903.

\bibitem{Barabaschi2023FEC}
Pietro Barabaschi, ``Progress on manufacturing, construction, commissioning and an updated baseline'', 29th Fusion Energy Conference, 16–21 Oct 2023, 2023, London, UK.

\bibitem{Klimov2013JNM}
N.S. Klimov, J. Linke, R.A. Pitts et al., ``Stainless steel performance under ITER-relevant mitigated disruption photonic heat loads'', J. Nucl. Mater., {\bf 438}, S241-S245 (2013);

\bibitem{Pitts2015JNM}
R.A. Pitts, B. Bazylev, J. Linke et al., ``Final case for a stainless steel diagnostic first wall on ITER'' J. Nucl. Mater. {\bf 463} (2015) 748–752.

\bibitem{Coenen2011PS}
J W Coenen, V Philipps, S Brezinsek et al., ``Analysis of structural changes and high-heat-flux tests on pre-damaged tungsten from tokamak melt experiments'' Phys. Scr. {\bf T145} (2011) 014066.

\bibitem{vanEden2014NF}
G.G. van Eden, T.W. Morgan, H.J. van der Meiden et al., ``The effect of high-flux H plasma exposure with simultaneous transient heat loads on tungsten surface damage and power handling'' Nucl. Fusion {\bf 54} (2014) 123010.

\bibitem{Yuan2016NF}
Y. Yuan, J. Du, M. Wirtz et al., ``Surface damage and structure evolution of recrystallized tungsten exposed to ELM-like transient loads'' Nucl. Fusion {\bf 56} (2016) 036021.

\bibitem{Wang2018NF}
Jun Wang, Chun Li, Yue Yuan et al., ``Surface modification of W–V alloy exposed to high heat flux helium neutral beams'' Nucl. Fusion {\bf 58} (2018) 096001.

\bibitem{Yuan2019NF}
Yue Yuan, Wangguo Guo, Peng Wang et al., ``Influence of surface melting on the deuterium retention in pure and lanthanum oxide doped tungsten'' Nucl. Fusion {\bf 59} (2019) 016022.

\bibitem{Hoelzl2021NF}
M. Hoelzl, G.T.A. Huijsmans, S. Pamela et al., ``The JOREK non-linear extended MHD code and applications to large-scale instabilities and their control in magnetically confined fusion plasmas'' Nucl. Fusion {\bf 61} 065001 (2021).

\bibitem{Di2021PPCF}
D Hu, G T A Huijsmans, E Nardon et al., ``Collisional-radiative non-equilibrium impurity treatment for JOREK simulations'' Plasma Phys. Control. Fusion {\bf 63} 125003 (2021).

\bibitem{Riccardo2005NF}
V. Riccardo, A. Loarte and the JET EFDA Contributors, ``Timescale and magnitude of plasma thermal energy loss before and during disruptions in JET'', Nucl. Fusion {\bf 45} 1427 (2005);

\bibitem{Rozhansky1995PPCF}
V Rozhansky, I Veselova and S Voskoboynikov, ``Evolution and stratification of a plasma cloud surrounding a pellet'', Plasma Phys. Control. Fusion {\bf 37} 399 (1995).

\bibitem{Matsuyama2022PRL}
A. Matsuyama, R. Sakamoto, R. Yasuhara et al., ``Enhanced Material Assimilation in a Toroidal Plasma Using Mixed H2 + Ne Pellet Injection and Implications to ITER'' Phys. Rev. Lett. 129, 255001 (2022).

\bibitem{CarrRaySect}
M. Carr et al., Raysect Python Raytracing Package (v0.4.0). Zenodo. https://doi.org/10.5281/zenodo.1205064.

\bibitem{CarrCHERAB}
M. Carr et al., cherab/core : Zenodo. https://doi.org/10.5281/zenodo.3551871

\bibitem{Kim2018NF}
S.H. Kim, T.A. Casper and J.A. Snipes, ``Investigation of key parameters for the development of reliable ITER baseline operation scenarios using CORSICA'', Nucl. Fusion {\bf 58} 056013 (2018);

\bibitem{Villone2010NF}
F. Villone, Y.Q. Liu, G. Rubinacci et al., ``Effects of thick blanket modules on the resistive wall modes stability in ITER'', Nucl. Fusion {\bf 50} 125011 (2010);

\bibitem{Zhang2020NF}
J. Zhang and P. B. Parks, ``Analytical formula for pellet fuel source density in toroidal plasma configurations based on an areal deposition model'', Nucl. Fusion {\bf 60} 066027 (2020);

\bibitem{Braginskii1965RPP}
S. I. Braginskii, ``Transport Processes in a Plasma'', Rev. Plasma Phys. {\bf 1}, 205, (1965);

\bibitem{ITERProgress}
E.J. Doyle, W.A. Houlberg, Y. Kamada et al., ``ITER Progress Chapter 2: Plasma confinement and transport'' Nucl. Fusion {\bf 47} S18–S127 (2007).

\bibitem{SpitzerBook}
L. Spitzer, Physics of Fully Ionized Gases (Interscience, New York, 1956).

\bibitem{Hirshman1978POF}
S.P. Hirshman, ``Neoclassical current in a toroidally‐confined multispecies plasma'', Phys. Fluids {\bf 21}, 1295 (1978).

\bibitem{ParkDistribution}
P. B. Parks, ``Modeling dynamics Fracture of Cryogenic Pellets'', GA report GA-A28352, (2016).

\bibitem{Gebhart2020FST}
T.E. Gebhart, L.R. Baylor \& S.J. Meitner, ``Shatter Thresholds and Fragment Size Distributions of Deuterium-Neon Mixture Cryogenic Pellets for Tokamak Thermal Mitigation'', Fusion Sci. Technol., 76:7 831-835 (2020).

\bibitem{Gebhart2021FST}
T. E. Gebhart, L. R. Baylor, and S. J. Meitner, ``Analysis of the Shattered Pellet Injection Fragment Plumes Generated by Machine Specific Shatter Tube Designs'', Fusion Sci. Technol., {\bf 77:1}, 33-41 (2021);



\bibitem{Aleynikov2024NF}
Pavel Aleynikov, Alistair M. Arnold, Boris N. Breizman et al., ``Thermal quench induced by a composite pellet-produced plasmoid'', Nucl. Fusion {\bf 64} 016009 (2024);

\end{thebibliography}
\end{document}